\documentclass{aa}  
\usepackage{graphicx}
\usepackage{xcolor}
\usepackage{scalefnt}
\usepackage[varg]{txfonts}
\def\Teff  {$T_\mathrm{eff}$}
\def\logg  {$\log g$}
\def\loggf {$\log gf$}
\def\vt    {$\rm v_{t}$}
\def\kms   {$\rm km\,s^{-1}$}

\begin{document}
\title{
Trans-iron Ge, As, Se, and heavier elements in the dwarf metal-poor stars \\
HD~19445, HD~84937, HD~94028, HD~140283, and HD~160617 \thanks{Based on
ultraviolet spectral observations made with the NASA/ESA Hubble 
Space Telescope, and 
obtained from the data archive at the Space Telescope Science Institute. 
STSci is operated by the Association of Universities for Research in 
Astronomy, Inc. under NASA contract NAS 5-26555. 
Based on optical spectral observations made primarily 
at the European
Organisation for Astronomical Research in the Southern
Hemisphere under ESO programs 
065.L-0507(A), 067.D-0439(A), 
068.D-0094(A) (PI Primas); 066.C-0220(A) (PI Sarre); 072.B-0585(A), 
074.B-0639(A) (PI Silva); 076.B-0133(A) (PI Pasquini); 
0101.A-0229(A) (PI Spite); and 266.D-5655(A); 
along with a spectrum from program 11AB01 (PI Barbuy) 
from the Canada-France-Hawaii 
Telescope (CFHT), which is operated by the National Research Council 
(NRC) of Canada, the Institut National des Sciences de l'Univers of the 
Centre National de la Recherche Scientifique (CNRS) of France, and the 
University of Hawaii; and a spectrum acquired at the Télescope Bernard Lyot
(TBL) at Pic du Midi Observatory with the spectrograph NARVAL under 
program L172N04 (PI Spite).
}}
%\fnmsep068
%\thanks{toto}
%}
\author{
[0000-0002-6775-9770]R.~C.\ Peterson\inst{1}
\and 
[0000-0001-9264-4417]B. Barbuy\inst{2}
\and
[0000-0002-2795-3421]M. Spite\inst{3}
 }
\institute {
SETI Institute, 189 Bernardo Ave. Suite 200, Mountain View, CA 94043, 
USA peterson@ucolick.org 
\and
Universidade de S\~ao Paulo, IAG, Rua do Mat\~ao 1226, 
Cidade Universit\'aria, 05508-900, S\~ao Paulo, Brazil
\and
GEPI, Observatoire de Paris, PSL Research University, CNRS,
Place Jules Janssen, 92190 Meudon, France
}
\date{Received ; accepted }
\authorrunning{Peterson et al.}
\titlerunning{Trans-iron elements in dwarf metal-poor stars}
\abstract
{
Spectra of unevolved metal-poor halo stars uniquely reflect the
elemental abundances incorporated during the earliest Galactic epoch.
Their heavy-element content is well understood as the products of neutron
capture on iron-peak elements. However, the lightest elements just past
the iron peak, with atomic number 30<Z<52, show striking abundance
patterns open to several interpretations. Understanding their nature may
illuminate the diverse halo, thick disk, or extragalactic origins of
metal-poor stars.

For five metal-poor dwarfs, we analyzed high-resolution echelle
UV spectra from the Hubble Space Telescope Imaging Spectrograph,
as well as archival optical echelle spectra. The goal
was to derive reliable halo dwarf abundances and
uncertainties for six trans-iron elements from UV spectra, and
optical abundances for four additional trans-Fe elements and two
well-understood heavier elements.

Our two independent analyses showed that the
largest source of discrepancy is UV continuum placement. Once
rectified, the internal results agree to 0.2 dex for moderately-unblended,
moderately strong lines. Our results similarly agree with previous work,
except where new data and unidentified Fe I lines are important. We
show that these heavily congregate
blueward of 2000A and redward of 2600A. Our
exclusion of trans-Fe lines blended by such lines
proved critical for arsenic.

A metallicity-dependent odd-even effect is uncovered among
trans-Fe elements: an odd-Z element abundance is depressed relative to
those of adjacent even-Z elements, especially at low metallicity. This is
supported by previous studies of Sr-Y-Zr, and also appears in some
theoretical calculations. To date, no calculations predict the
high Mo/Ge ratio, independent of Mo/Fe, that we find in all five stars.
Our work thus highlights the complexity of
the production of the light trans-Fe elements in metal-poor
stars.
}
\maketitle
%
%---------------------------- Introduction -----------------------
\section{Introduction}

The assembly sequence of our Galactic halo is encoded in spectra of
its surviving low-metallicity turnoff stars, the unevolved
stellar survivors from the earliest epochs whose elemental abundance
ratios reflect the nucleosynthesis processes and the nature of the
stars creating them.  As reviewed by \citet{SnedenCG08}, 
most halo dwarfs whose iron abundances Fe/H\footnote{Abundance definitions:
Iron abundance Fe/H = $N_{Fe} / N_{H}$.\\
$\rm [Fe/H]=  log(N_{Fe} / N_{H})_{star} - log(N_{Fe} / N_{H})_{Sun}$.\\
For element X, $\rm [X/Fe]=  log(N_{X} / N_{Fe})_{star} - log(N_{X} / N_{Fe})_{Sun}$.\\
A(X)= log[(N(X)/N(H)] + 12.}
are 1/30 to 1/1000 
that of the Sun ($-3.0 \leq$ [Fe/H] $\leq -1.5$) 
show modest abundance enhancements of light-alpha
elements. This overabundance suggests that these elements 
originate in supernovae evolving from one or more
massive stars \citep[e.g.,][]{WoosleyW95}, exploding shortly after the Big Bang. 
The rare stars with even greater iron deficiencies often have extreme overabundances of carbon, 
nitrogen, and oxygen, and generally magnesium as well; 
these carbon-enhanced metal-poor (CEMP) stars undoubtedly require either additional internal processing
or the products of a second massive-star explosion 
\citep[e.g.,][]{LimongiCB03}.

The production of the elements heavier than the iron peak, 
in both metal-poor and more metal-rich stars, 
is also well understood as being due to 
the s-process and the r-process, the slow and rapid capture 
of neutrons by iron-peak elements. 
As first noted by \citet{Truran81}, unevolved moderately metal-poor 
stars generally lack the s-process contribution; their  
entire heavy-element content is better explained by 
the r-process, possibly in a single prior event. 
As more results for 
such stars became available, the overall r-process content was 
found to vary among them by several orders of magnitude 
\citep[see e.g.,][]{FrancoisDH07}. 
\citet{KratzFP07} and \citet{SnedenCG08} emphasize and illustrate 
that regardless of the overall level of r-process abundances,
the relative proportion of the heaviest r-process elements 
always remains the same and matches that of the Solar System, 
especially among those elements of the second peak 
between Ba at Z=56 and Pt at Z=78. This remains true even among CEMP stars, 
most of which are giants; many also have sizable levels 
of r-process and/or 
%or 
s-process enhancement. Likewise, in 
extremely metal-poor giants with more normal CNO abundance levels, 
the ratio Eu/Ba remains
constant whatever the abundance level of the r-process elements
\citep[e.g.,][]{MashonkinaCB10,SpiteSB18}. 
However, as \citet{KratzFP07} note, the lighter elements just past the 
iron peak are not so well behaved, and lead to suggestions for 
"a second distinct (weak) r-process site" for these elements. 
Extensive attempts to pinpoint such a process or processes 
\citep[e.g.,][]{Wanajo07,Wanajo13} have yet to definitively reveal a solution.

The great age of very metal-poor stars (as indicated by their 
space motions and nuclear chronometers) implies that the r-process has operated 
at the earliest times of Galactic history, as expected if 
it arises in massive-star explosions. In contrast, 
the s-process is believed to occur only in evolved stars, 
especially those on the asymptotic 
giant branch (AGB). Thus s-process production is delayed by the time it 
takes stars to evolve from the main sequence, which rapidly 
increases as mass decreases. Understanding 
these time scales could establish the sequence of events leading to 
the appearance of s-process contributions, which in turn might 
shed light on the diverse halo, thick disk, or extragalactic 
origins of the metal-poor stars in our Galaxy. 

However, separating the s-process and r-process contributions in 
any given star is difficult. Only for the s-process are synthesis 
yields predicted with reasonable reliability, as they can be derived 
from neutron-capture cross sections. 
This is best done for the Sun and solar system
abundances, where isotopic abundances are
available \citep[e.g.,][]{Lodders10}.
The r-process contribution to the solar abundances is often derived 
empirically by calculating and subtracting the expected s-process 
contribution for each isotope \citep[e.g.,][]{SimmererSC04}. 
But at the solar age 
of a few billion years, there is no guarantee that all heavy elements 
were formed exclusively by a single common r-process, nor in a single event. 
At much earlier times, different environments and processes may have dominated. 

Their discernment 
might come from abundance proportions of elements with intermediate 
atomic number $30<Z<52$, the trans-Fe elements just beyond iron. 
As noted above, their production is not nearly as well understood, and 
the various scenarios proposed have widely varying time scales. 
Examples are an i-process with intermediate neutron flux 
\citep[e.g.,][]{CowanRose77,RoedererKP16}, a Light Element Primary 
Process \citep[LEPP; ][]{TravaglioGA04},
the low-entropy domain of the neutrino-driven
winds above the core of a collapsing Type II supernovae
\citep{FarouqiKP09,FarouqiKM09,ArconesMontes11} or 
a ``weak'' r-process in the wind above the core of electron-capture supernovae
\citep{Wanajo13,NiuCZ15}, 
pulsating evolved AGB stars \citep{BisterzoTG14}, 
or fast-rotating main-sequence stars \citep{FrischknechtHP16}.

The calculations required for reliable yield predictions for trans-Fe 
and heavier elements are very complex. Only recently have trans-Fe 
yields become available for  
stars over the range of low metallicities that characterize the 
Galactic halo.
For example, \citet{BanerjeeQH17}  
provided yield patterns from Ge (Z=32) to Pb (Z=83) that result from various 
degrees of ingestion of surface hydrogen into a helium-burning shell 
of an evolved massive star of 25M\sun, over metallicities ranging 
from one-tenth to 1/10,000 solar. 
\citet{RitterHJ18} have extended their NuGrid model predictions 
to the metallicities of halo stars near the main-sequence turnoff, 
providing stellar overproduction factors for virtually 
all elements synthesized in stars with initial masses 1 -- 25 M\sun\ 
as they evolve, and incorporating the neutron addition processes 
from the neutron fluxes as they arise.

Observational constraints for trans-Fe elements are also increasing, but were originally confined to metal-poor giants. Lines of 
trans-Fe elements are typically weak in the spectra of metal-poor stars, 
due to their modest abundances, and are generally 
detectable only in the crowded ultraviolet spectral region. Notable 
exceptions are the well-observed trans-Fe elements strontium (Z=38), 
yttrium (Z=39), and zirconium (Z=40), which have 
many optical lines. Germanium (Z=32) and molybdenum (Z=42) 
each have a single line potentially detectable from the ground, 
but only in cooler metal-poor stars. 

With the advent of the Space Telescope Imaging Spectrograph (STIS) on 
the Hubble Space Telescope, panchromatic high-resolution UV spectra 
have become feasible for the brightest metal-poor stars. Such spectra 
enabled the unanticipated \citet{Peterson11} discovery that in 
several halo turnoff metal-poor stars, all with r-process enhancements 
of less than a factor of four, the molybdenum-to-iron abundance 
ratio is six to ten times that of the Sun. As \citet{Peterson11} 
noted, this contradicts previous results for metal-poor giants, 
where the single available optical line indicates a near-solar 
Mo/Fe abundance ratio for all halo and globular cluster giants 
except those with more extreme r-process enhancements. 

Further UV work on metal-poor stars has revealed that many 
exhibit remarkable star-to-star variations among the relative 
abundances of several of these trans-Fe elements, in contrast to 
the internal conformity of the heavier elements within both the 
s-process and the r-process. 
\citet{Peterson11,Peterson13} and \citet{RoedererKP16} saw striking variations
among metal-poor turnoff stars, as did
\citet{SiqueiraSB13} among metal-poor giants. Addressing this again 
using spectra redward of 3500\AA, \citet{SpiteSB18} uncovered trans-Fe 
abundance variations among a dozen giants (including one CEMP star) 
with iron abundances $<$ 1/300 solar, 
whose r-process enhancements ranged from 
solar to ten times solar.

The aim of the present work is to address a single group of low-metallicity,
low-r-process, unevolved stars as homogenously as possible. 
Our goal in this work is to
analyze spectra obtained with the Space Telescope Imaging
Spectrograph (STIS), primarily with the highest resolution grating E230H,
that can isolate enough lines of trans-Fe elements
to provide abundances sufficiently
reliable to distinguish the various nucleosynthesis
mechanisms operational at the very earliest times. To extend the
analysis to heavier elements, high-resolution optical spectra are analyzed
as well.

From them we have determined abundances and constrained their uncertainties for a range of
elements within and beyond the trans-Fe domain in the five
metal-poor turnoff stars HD\,19445, HD\,84937, HD\,94028,
  HD\,140283, and HD\,160617. %\\
In Sec.\ 2, several recently suggested sites of trans-Fe element synthesis are reviewed.
In Sec.\ 3, the observations adopted here are reported.
Section 4 summarizes the abundance analysis methods.
Section 5 presents plots of calculations versus observations that
illustrate the determination of trans-Fe elemental abundances. 
It characterizes the presence of newly identified {\ion{Fe}{I}} lines 
and lines that remain unidentified, and shows their 
influence on certain trans-Fe lines often adopted previously.
Abundance results and their constraints on i-process calculations 
are discussed in Sec.\ 6.  A summary is provided in Sec.\ 7.

\section{Recent suggestions for the sites of heavy-element synthesis in metal-poor stars}

As \citet{CowanSL19} recently reviewed, the 
astrophysical site or sites of the r-process may include more exotic 
environments than the collapse and explosion of a single massive star, 
and might lead to diverse yields. 
One possibility is the merger of two neutron stars, encouraged by 
recent observations of the kilonova associated with gravitational wave
event GW170817 \citep{AbbottAA17,AbbottAAA17}. 
Such events are capable of 
r-process nucleosynthesis that produces heavy
elements \citep[notably strontium:][]{WatsonHS2019}.
Other scenarios include
the merger of a neutron-star with a black hole, 
rarer classes of massive single-star supernovae,
magneto-rotational supernovae with jets \citep{WintelerKP12}, 
and collapsars \citep{SiegelBM18}.
Like a massive, normally evolving single-star supernova, all these more exotic 
explosive and merger events take place very shortly after
star formation first occurs, on the time scale of the evolution
of massive stars. Given the complexity of the energy and dynamics of 
such explosions, detailed predictions of heavy-element 
production are difficult to provide. 

Several tens of Myr later, a contribution can also occur 
from fast rotating
massive stars (FRMS). This requires about 40 Myr for stars 
of seven solar masses 
\citep[][]{MeynetEM06,HirschiMM07,FrischknechtHP16}.
A few hundred Myr later, 
contributions may be incorporated from evolved
low-mass stars ($\leq 2$ solar masses)
on the asymptotic giant branch (AGB) or super AGB (SAGB)
\citep{GallinoBS06,KappelerGB11,BisterzoTW17}.
Both these scenarios may also produce heavy elements by the s-process, 
and under the right circumstances can also synthesize certain 
lighter elements.

%TABLE 1
\begin{table*}
\label{loghst}      % is used to refer this table in the text
%\scalefont{1.0}
%Fails: \captionsetup{justification=centering}  \parbox: Caption is centered but multilined below
%\parbox{4cm}{\caption{Log of adopted HST STIS E230H and E230M echelle spectra}}
\caption{Log of adopted HST STIS E230H and E230M echelle spectra}
\centering                          % used for centering table
\begin{tabular}{ccccccccc}        % centered columns (4 columns)
\hline%\hline                 % inserts double horizontal lines
\noalign{\smallskip}
\hbox{Star} & Program ID & Instrument & PI & \hbox{Date} & Central wavelengths \\
            &            &            &    &             &   ({\rm \AA})       \\
%\noalign{\smallskip}
\hline
\noalign{\smallskip}
HD 19445 & 14672      & STIS  E230H & Peterson & 2017  & 2013 & \\
         & 7402       & STIS  E230M & Peterson & 1999  & 2707 & \\  
\hline
\noalign{\smallskip}
HD 84937 & 14161      & STIS  E230H & Peterson & 2016  & 2013, 2263, 2513, 2762, 3012 \\
%        & 7402       & STIS  E230M & Peterson & 2000  & 2707 & \\     
\hline
\noalign{\smallskip}
HD 94028 & 8197       & STIS  E230H & Duncan   & 2000  & 2013 & \\
         & 14161      & STIS  E230H & Peterson & 2016  & 2263, 2513, 2762, 3012 & \\
%        & 7402       & STIS  E230M & Peterson & 1998  & 2707 & \\     
\hline
\noalign{\smallskip}
HD 140283& 7348       & STIS  E230H & Edvardsson& 1999  & 2063, 2113, 2163 & \\
         & 9455       & STIS  E230H & Peterson & 2002  & 2513, 2762, 3012 & \\
         & 9491       & STIS  E230H & Balachandran&2003& 3012 & \\
         & 14161      & STIS  E230H & Peterson & 2016  & 2263, 2513, 2762 & \\
         & 14672      & STIS  E230H & Peterson & 2017  & 1963 & \\
\hline
\noalign{\smallskip}
HD 160617& 8197       & STIS  E230H & Duncan   & 1999-2000  & 2013 & \\ 
%\noalign{\smallskip}
\hline                                   %inserts single line
\end{tabular}
%\caption{Log of adopted HST STIS E230H and E230M echelle spectra\label{loghst}}
\end{table*}

Results from UV echelle spectra for lighter neutron-capture 
elements, such as Ge, As, Se, Mo, and Ru, have led 
 various authors to weigh in on the 
implications for trans-Fe synthesis sites at early times.
For five warm, metal-poor turnoff stars, none of which had 
an r-process content 
enhanced by as much as a factor of four above solar, \citet{Peterson11} 
found that all five 
showed elevated abundances of Mo and Ru with respect to iron, but 
that the somewhat heavier element Cd (Z=48) and the s-process
element La (Z=57) were minimally enhanced. This suggested the operation 
of high-entropy winds (HEW) operating above the core of collapsing
Type II supernovae \citep{FarouqiKP09,FarouqiKM09}, the only mechanism
that predicted the production of light trans-iron elements confined to a
narrow mass range. 

In a sample of six metal-poor stars, including three that \citet{Peterson11} had analyzed, \citet{Roederer12} studied the abundance of Ge,  As, and Se, 
and suggested that the weak and main components of the s-process are not 
likely to be the sources of these three elements.  

Turning to very metal-poor giants, and incorporating optical spectra to extend 
the comparisons to many more heavy elements, \citet{SiqueiraSB13} compared the abundance
distributions of the both the trans-Fe and the heavier elements in HD~122563 (with low heavy
r-process content) versus those in CS-31082-001 (with high heavy r-process content). They
suggested that the light neutron-capture elements can be formed in
electron-capture supernovae as well \citep{WanajoJM11}.%\\

\citet{RoedererKP16} found a high ratio of the abundance 
of arsenic relative to those of Ge and Se for the mildly metal-poor star 
HD~94028, whose composition shows s-process as well as r-process products. 
To explain this they suggested an i-process like that of 
\citet{CowanRose77}, in which synthesis occurs at a neutron flux intermediate 
between those of the s- and r-processes, and inferred it had operated 
in early epochs. 

Finally \citet{SpiteSB18} showed from optical spectra 
that, in extremely metal-poor giants,
the behavior of the light neutron-capture elements between Sr (Z=38) 
and Ag (Z=47), whose proportions differ between 
r-rich and r-poor stars, can only be explained
by a heavy element enrichment in two steps. Several authors had also 
suggested this previously for trans-Fe elements in metal-poor stars 
\citep[e.g.,][]{CescuttiCH13,HansenMA14,Roederer17}. 
\citet{SpiteSB18} proposed an 
enrichment in main r-process elements that would lead to the abundance
pattern observed in the classical r-rich stars (like CS\,31082-001),
and that, independently, a second mechanism detectable only in r-poor
stars would enrich the matter in mainly first peak elements.  
Both processes would be required to operate at very early 
galactic epochs, since extremely metal-poor giants are believed to 
be very old.\\

\section {Observational data}
\label{uvobs}

The UV analysis of this work is primarily based on HST spectra acquired with the
E230H grating of the STIS echelle spectrograph and the 0.2" x 0.09"
slit for a resolving power $R$=114\,000. 
The log of observations with the STIS spectrograph is given in
% 2020/01/18 v25ruth3   Otherwise loghst table is referenced as Table 2
Table 1. %\ref{loghst}.

 The E230H spectra provide complete high resolution coverage of the
 stellar spectrum in the interval 1880-3140 {\rm \AA} for HD\,84937,
 HD\,94028, and HD\,140283.  For HD\,19445, E230H coverage is 
 limited to 1880-2150 {\rm \AA}; E230M spectra are
 available between 2280 and 3179 {\rm \AA}.  For HD\,160617, E230H 
 spectra again cover 1880-2130\AA, beyond which no 
 UV STIS echelle spectra are available.  At 
 2000 {\rm \AA}, the E230H spectra all show S/N $>$ 50.

We have simultaneously re-analyzed high-resolution optical archival
spectra 
to obtain abundances of the first peak elements Sr, Y, Zr, and
the second peak elements Ba and Eu.  For those elements where both
optical and UV lines are useful, this allows a direct comparison
of the results of the more widely used optical spectra with those
based on UV spectra.  The optical spectra adopted are described
below.%\\

%Table 2
\begin{table*}
\begin{center}
%Fails: \captionsetup{justification=centering}   \parbox: Caption is centered but multilined below
%\parbox{4cm}{\caption[]{Stellar Parameters adopted with {\tt Turbospectrum} and with {\tt SYNTHE}}}
%Below also fails to override default caption indent
\hangindent=1.5cm{\caption[]{Stellar Parameters adopted with {\tt Turbospectrum} and with {\tt SYNTHE}}}
\label{param}
\begin{tabular}{rccrrcrrrrrrr}
\hline
\noalign{\vskip 0.05cm}
%\cline{3-7} \cline{11-13} 
         \hbox{ } &~~~|~~~&  \multispan5 \hfill  {\tt Turbospectrum} \hfill
         &~~~~~~~~~~~~~|~~~~~~~~~~~~~& \multispan5 \hfill  SYNTHE \hfill  \\
\hbox{Star}     &~~~|~~~& \Teff &\logg & [Fe/H] &\hbox{\vt } &Ref &~~~~~~~~~~~~~|~~~~~~~~~~~~~&\Teff &\logg & \hbox{[Fe/H]} & \hbox{\vt } & Ref\\
\noalign{\vskip 0.05cm}
\noalign{\vskip 0.05cm}
\noalign{\hrule}
\noalign{\vskip 0.05cm}
%\rule{0pt}{2ex}
\hbox{HD\,19445}  &~~~|~~~& 6070 &  4.4 & -2.15 & 1.3 &1&~~~~~~~~~~~~~|~~~~~~~~~~~~~&6050 & 4.5 & -2.00 & 1.0 & 6 \\
\hbox{HD\,84937}  &~~~|~~~& 6300 &  4.0 & -2.25 & 1.3 &2&~~~~~~~~~~~~~|~~~~~~~~~~~~~&6300 & 4.0 & -2.25 & 1.3 & 3 \\
\hbox{HD\,94028}  &~~~|~~~& 6050 &  4.3 & -1.40 & 1.2 &3&~~~~~~~~~~~~~|~~~~~~~~~~~~~&6050 & 4.3 & -1.40 & 1.2 & 3 \\
\hbox{HD\,140283} &~~~|~~~& 5750 &  3.7 & -2.57 & 1.4 &4&~~~~~~~~~~~~~|~~~~~~~~~~~~~&5700 & 3.6 & -2.60 & 1.3 & 3 \\
\hbox{HD\,160617} &~~~|~~~& 5950 &  3.9 & -1.80 & 1.3 &5&~~~~~~~~~~~~~|~~~~~~~~~~~~~&6000 & 3.8 & -1.80 & 1.2 & 6 \\
\hline
%\noalign{\vskip 0.05cm}
%\noalign{\vskip 0.05cm}
%% Use \tablefoot and \tablebib
%\multicolumn{12}{l} {Note: $\rm [Fe/H]=  log(N_{Fe} / N_{H})_{star} - log(N_{Fe} / N_{H})_{Sun}$} \\
%\multicolumn{12}{l} {References for model parameters: 1: \citet{RoedererHV18};~~2: \citet{SpitePG17};~~3: \citet{PetersonKA17};} \\
%\multicolumn{12}{l} {~~~~~~4: \citet{SiqueiraAB15};~~5: \citet{RoedererLawler12};~~6: \citet{PetersonK15}} \\
%\hline
\end{tabular}
\end{center}
\tablefoot{$\rm [Fe/H]=  log(N_{Fe} / N_{H})_{star} - log(N_{Fe} / N_{H})_{Sun}$}
%{\bf $\rm [Fe/H]=  log(N_{Fe} / N_{H})_{star} - log(N_{Fe} / N_{H})_{Sun}$}
%}
\tablebib{(1)~\citet{RoedererHV18}; (2) \citet{SpitePG17}; (3) \citet{PetersonKA17}; 
(4) \citet{SiqueiraAB15}; (5) \citet{RoedererLawler12}; (6) \citet{PetersonK15}}
%{\bf (1)~\citet{RoedererHV18}; (2) \citet{SpitePG17}; (3) \citet{PetersonKA17}; 
%(4) \citet{SiqueiraAB15}; (5) \citet{RoedererLawler12}; (6) \citet{PetersonK15}
%}
%}
%Notes. References for model parameters: 1: \citet{RoedererHV18};~~2: \citet{SpitePG17};~~3: \citet{PetersonKA17};
%~~4: \citet{SiqueiraAB15};~~5: \citet{RoedererLawler12};~~6: \citet{PetersonK15}.
%$\rm [X/H]=  log(N_{X} / N_{H})_{star} - log(N_{X} / N_{H})_{Sun}$.
\end{table*}

\section{Analysis methods}
\label{meth}

To investigate analytical uncertainties, two independent
approaches to the abundance analyses were followed.  Spite and Barbuy
performed a $\chi^{2}$ fit of the observed spectrum to a synthetic
spectrum computed with the spectrum synthesis code {\tt Turbospectrum}
\citep{AlvarezP98,Plez12}.  
Peterson ran the spectrum synthesis code {\tt SYNTHE} \citep{Kurucz93,Kurucz04} over the
region from 1850 {\rm \AA} to 6000 {\rm \AA}, deriving abundances
visually by altering the elemental abundances adopted for {\tt SYNTHE}
until the calculated strengths of the relevant lines of an element
matched those in the observed spectrum.  The {\tt SYNTHE} 
input line list is a more recent version 
of the one employed by \citet{PetersonKA17}, in which 
log gf-values are originally adopted from the Kurucz web site 
and continually updated to best match high-quality spectra for stars 
spanning a range of line strengths, as illustrated below.

Both {\tt Turbospectrum} and {\tt SYNTHE} include standard sources 
of continuous opacity. In the UV, both incorporate the bound-free 
edges of several abundant heavy elements whose strong influence 
is illustrated in Fig.\ 8.6 of \citet{Gray05}. That work emphasizes 
that heavy elements dominate the continuous opacity below 2500\AA, and  
increases so rapidly toward bluer wavelengths that the continuum optical 
depth is forced to progressively higher, shallower atmospheric levels. \citet{Gray05} 
also notes the importance of Rayleigh scattering in the UV in cool stars, 
which again is included in both {\tt Turbospectrum} and {\tt SYNTHE}. 

The opacity of overlapping individual lines also contributes 
at and below 2000A, and in redder regions wherever strong 
lines or line wings dominate. Consequently, even the weakest UV lines 
near 2000\AA\ are formed at shallow atmospheric levels. 

The same model atmospheres were adopted for the optical analysis as
for the UV. Models for the {\tt Turbospectrum} analysis were interpolated
in the MARCS grids \citep{GustafssonEE08}. 
The {\tt SYNTHE} analysis adopted interpolations of models from the grids of
\citet{CastelliKur03}.  For the extremely transparent atmospheres of
the weak-lined stars HD~84937 and HD~140283, the {\tt SYNTHE} analysis 
adopted models interpolated from 
the versions subsequently made publicly available by Fiorella Castelli
that assume a mixing-length to scale-height ratio of 0.5, instead of
the standard 1.25 value of the other models.  

With these choices, {\tt SYNTHE} calculations for the all five stars 
yielded no discernable
discrepancy in the abundances of {\ion{Fe}{I}} lines regardless of wavelength
or lower excitation.  This eliminates a tendency that persists in some
current work \citep[see e.g.,][ and works cited therein]{RoedererHV18}
for the iron abundances found from {\ion{Fe}{I}} lines of lowest excitation to
diverge from those found from {\ion{Fe}{II}} lines and {\ion{Fe}{I}} lines of higher
excitation.

Both {\tt Turbospec} and {\tt SYNTHE} analyses adopted the high-resolution UV
% 2020/01/18 v25ruth3   Otherwise loghst table is referenced as Table 2
spectra in Table 1. %\ref{loghst} 
%for their UV analysis.
However, in the optical, the analyses incorporated
different observational spectra.  For the optical {\tt Turbospectrum}
analysis, ESO spectra obtained with the Very Large Telescope (VLT) and
the spectrograph UVES \citep{DekkerDK00} were adopted.  Included were
the observations in the UVES programs listed in the footnote on the first page. 
%065.L-0507(A), 067.D-0439(A),
%068.D-0094(A) (PI Primas); 066.C-0220(A) (PI Sarre); 072.B-0585(A),
%074.B-0639(A) (PI Silva); 076.B-0133(A) (PI Pasquini); 0101.A-0229(A)
%(PI Spite); 266.D-5655(A).  
In all cases, the resolving power $R$ was
47\,000 or higher.  These spectra were reduced using the UVES context
\citep{BallesterMB00}.  The S/N of the resulting spectra exceeded 100
at 350nm.  To better detect extremely weak lines, this analysis also
incorporated a high S/N spectrum of HD\,140283 obtained with the
spectrograph ESPADONS at the CFHT in Hawaii under program 11AB01 
(PI Barbuy). This had a resolving power
$R$ = 81\,000, and covered the wavelength range 3700-10475 {\rm \AA}
with S/N = 800-3400 per pixel.

%\\

The {\tt SYNTHE} optical analysis also relied on UVES archival
spectra, along with Keck HIRES spectra from the KOA archive.  Table 1
of \citet{PetersonK15} includes the specifics of the optical spectra
adopted for HD~94028, HD~140283, and HD~160617 from KOA.  While these 
supported the {\tt Turbospectrum} analysis, they are less accurate, 
being based on spectra of lower S/N and resolution. Consequently the 
{\tt SYNTHE} optical values were not included in the final results below.

\subsection{Stellar Parameters}
\label{stpar}

The stellar parameters adopted for the five stars in each approach are
% 2020/01/18 v25ruth3   Otherwise param table is referenced as Table 3
reported in Table 2 %\ref{param} 
along with their sources.
The {\tt SYNTHE} parameters were derived strictly spectroscopically
\citep{PetersonDR01,PetersonKA17}, 
while the {\tt Turbospectrum} parameters also incorporate photometry. 

Despite their different origins, the parameters adopted for the 
{\tt Turbospectrum} and {\tt SYNTHE}
computations are very similar.  The difference in the adopted values is
at most 50\,K for the effective temperature \Teff, 0.1\,dex in gravity
\logg, 0.3\,\kms\ in microturbulent velocity \vt, and 0.15\,dex in
[Fe/H]. 
Runs with {\tt Turbospectrum} adopting the {\tt SYNTHE} models 
confirmed that these small differences in the model parameters 
have a negligible influence on the 
abundances deduced. 

Our temperature for HD\,94028 is 300K higher than the values of 
\citet{Roederer12} and \citet{RoedererKP16}. However, 
\citet{RoedererSL18} redetermined the stellar parameters of HD\,94028,
and also HD\,19445 and HD\,84937, and these values 
of temperature, gravity and microturbulent velocity now agree 
% 2020/01/18 v25ruth3   Otherwise param table is referenced as Table 3
with our determinations in Table 2 %\ref{param} 
to within the errors.%\\

\subsection {Selection of lines and gf-values}
\label{linegfselect}

The list of lines used for the determination of the abundances of
the trans-iron and neutron-capture elements is given in Table
\ref{linelist}, together with the adopted log gf 
values. The latter are largely
taken from \citet{Morton00}.  Where these are unavailable or have been
revised based on improved laboratory measurements, other sources were
adopted, as specified in the table. As an example, for cadmium we adopted 
the updated \citet{RoedererLawler12} log gf value for the {\ion{Cd}{I}} line at 
2288\AA, as well as their hyperfine splitting for both cadmium lines, 
but reverted to the \citet{Warner68b} gf value for the 
{\ion{Cd}{II}} line at 2144\AA, because that value provided 
excellent consistency between the two lines in our abundance determinations. 
To illustrate the effect of the hyperfine splitting of this feature as lines become stronger, 
its hyperfine splitting was not included in the {\tt SYNTHE} calculations shown for this line in the next section. 

For each line, Table \ref{linelist} includes the abundance of the
element A(X) (defined as $\rm A(X) = log(N_{X} / N_{H}) + 12$). 
The set of A(X) values listed on the left were derived with {\tt Turbospec}; those to the right were derived with {\tt SYNTHE}. Both analyses adopted the line parameters appearing in the first five columns. 
Wavelengths for {\tt SYNTHE} calculations were adjusted by up to 
0.02\AA\ where necessary, to 
match the wavelength scale set by \ion{Fe}{I} \citep{PetersonKA17}.

We tried to avoid the lines too severely blended. 
This was judged both from the line list itself, and 
from the figures below comparing spectral synthesis versus
observation for each star. 

For germanium, many lines are potentially available, but all are weak in the most
metal-poor stars. We chose as many lines as feasible:
eight minimally blended  {\ion{Ge}{I}} lines with consistent experimental and
theoretical gf-values from \citet{LiNPW99}.
The {\ion{Ge}{I}} log gf values in Table \ref{linelist} are
the straight average of their theoretical and experimental results.

The {\tt SYNTHE} analysis yielded results for all eight {\ion{Ge}{I}} lines
in all five stars, except for those regions where data are lacking for HD~160617 (Table 1).
The low internal dispersion of the
{\tt SYNTHE} Ge abundances determined for each individual star seen in Table \ref{linelist} 
supports the internal consistency of the \citet{LiNPW99}
gf values for these eight lines.

As described below in discussing the figures, for arsenic, selenium, and germanium 
we excluded certain lines adopted in previous work that we found to be 
blended by {\ion{Fe}{I}} lines listed in the laboratory {\ion{Fe}{I}} study of \citet{BrownGJ88}, 
and those recently identified by \citet{PetersonK15} or \citet{PetersonKA17}. 
Consequently, abundances deduced for these elements tend to be lower 
in this versus previous work, especially for arsenic where few lines are available.

\subsection{Comparison of methods and results from each approach}
\label{intercomp}

Because the same log gf-values were adopted for the UV 
lines of trans-Fe species, the line-by-line differences in the 
abundances found by these 
independent analyses are useful in assessing the uncertainties 
in the derivation of the abundances of individual elements. 
In general, these agree to better than 0.1 dex, when based on largely
unblended lines of medium strength in regions where the continuum is
well defined.  Line-to-line differences reach 0.2 dex when 
lines become blended 
or very weak, and can exceed this both for saturated lines 
and for lines blueward of 2000\AA, where the S/N of the observed spectra decreases and 
the number of unidentified lines rises rapidly, as illustrated just below.

We also investigated potential UV systematics due strictly to the code
and the source of models, by incorporating both the line lists and 
the \citet{CastelliKur03}
models adopted by Peterson into {\tt Turbospec} runs. 
No differences
were discerned at a level of $~$0.05 dex 
\citep[see also Appendix A in ][]{BonifacioSC09}.

\section{Comparisons of observed and calculated UV spectra near lines of trans-Fe elements, and the effect of unidentified lines}
\label{uvfig}

\begin{figure}%[]
%\resizebox{8.0cm}{12.0cm}              
%{\includegraphics [scale=0.55]{figa190.pdf}}
%\centering
\resizebox{\hsize}{!}{\includegraphics{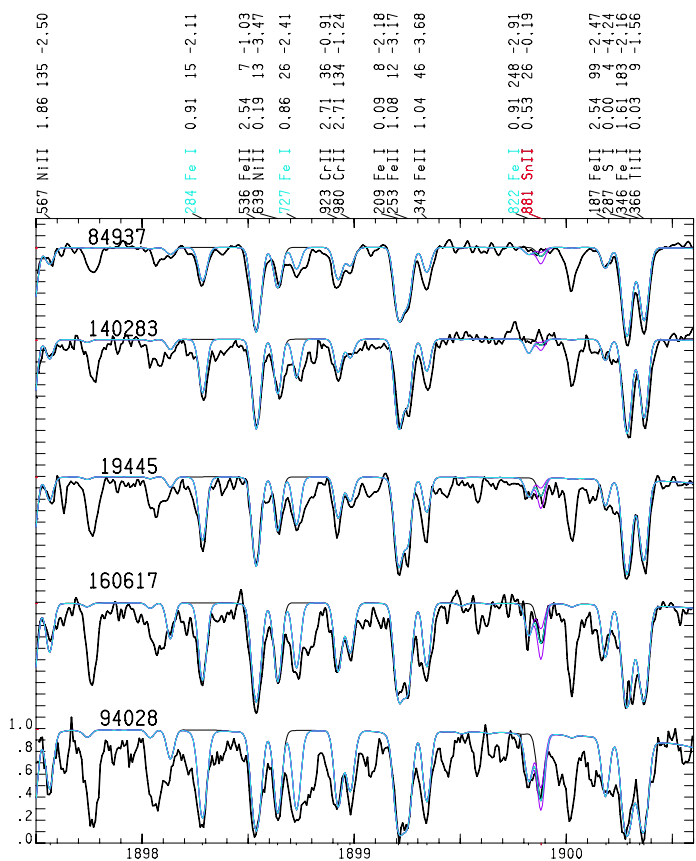}}
\caption{%[]
Spectra for the five stars are plotted in the region of the \ion{Sn}{II} 1899.898 {\rm \AA} line, where unidentified lines are numerous and strong. 
%{\bf 
Each of Figs.\ 1 -- 15 compares the observed spectrum (heavy black line) 
with the two calculations, one lacking newly identified {\ion{Fe}{I}} 
lines (thin black line) and one including them (light blue line). 
The Y scale on the left is for the bottom comparison. The others
have the same vertical scaling factor; ticks on the Y axis indicate
a displacement of 0.1 full scale. At the bottom is the wavelength
in {\rm \AA}.
At the top, aligned with line wavelength, are listed 
the line parameters for the strongest lines in the best-fit calculation
of the bottom spectrum. 
Following the decimal digits of the wavelength are
given the species, the lower excitation potential in eV, the decimal digits of the core 
residual intensity of the unbroadened spectrum, and the log gf-value adopted. 
Lines of trans-Fe species are highlighted in red, and those of newly identified {\ion{Fe}{I}} 
in light blue. Purple lines indicate light-blue calculations with a $\pm$0.3 change in
the adopted abundances of each trans-Fe element. 
}
\label{fig1}
\end{figure}

\begin{figure}%[p]
%\resizebox{8.0cm}{12.0cm}              
%\includegraphics[scale=0.45]{figa189.pdf}
{\includegraphics [scale=0.40]{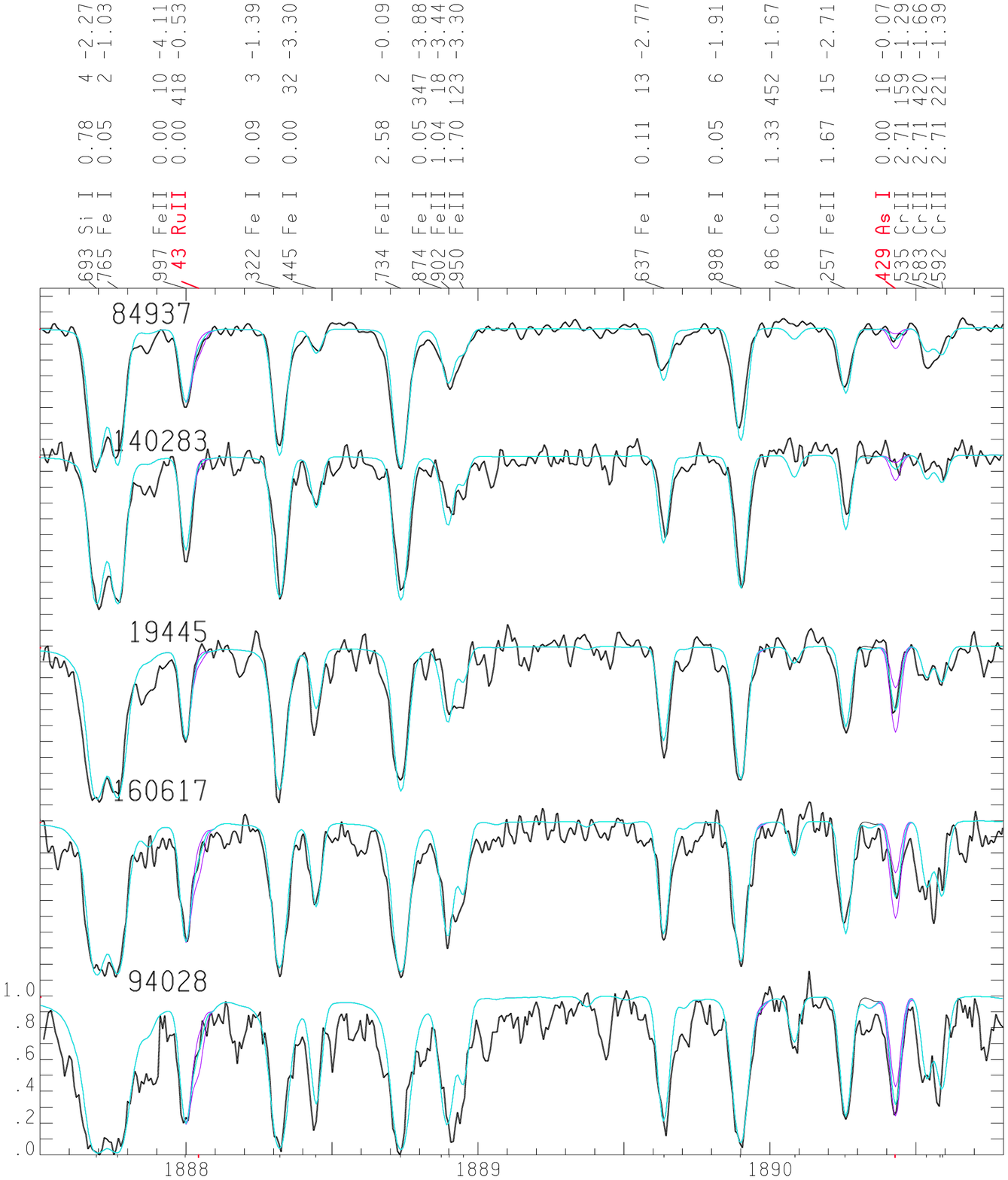}}
%\centering
%\resizebox{\hsize}{!}{\includegraphics{figb189.eps}}
\caption{Spectra of the \ion{Ru}{II} 1888.064 {\rm \AA} and \ion{As}{I} 1890.429 {\rm \AA} lines.}
\label{fig2}
\end{figure}

\begin{figure}%[p]
%\resizebox{14.0cm}{12.0cm}           	
{\includegraphics [scale=0.40]{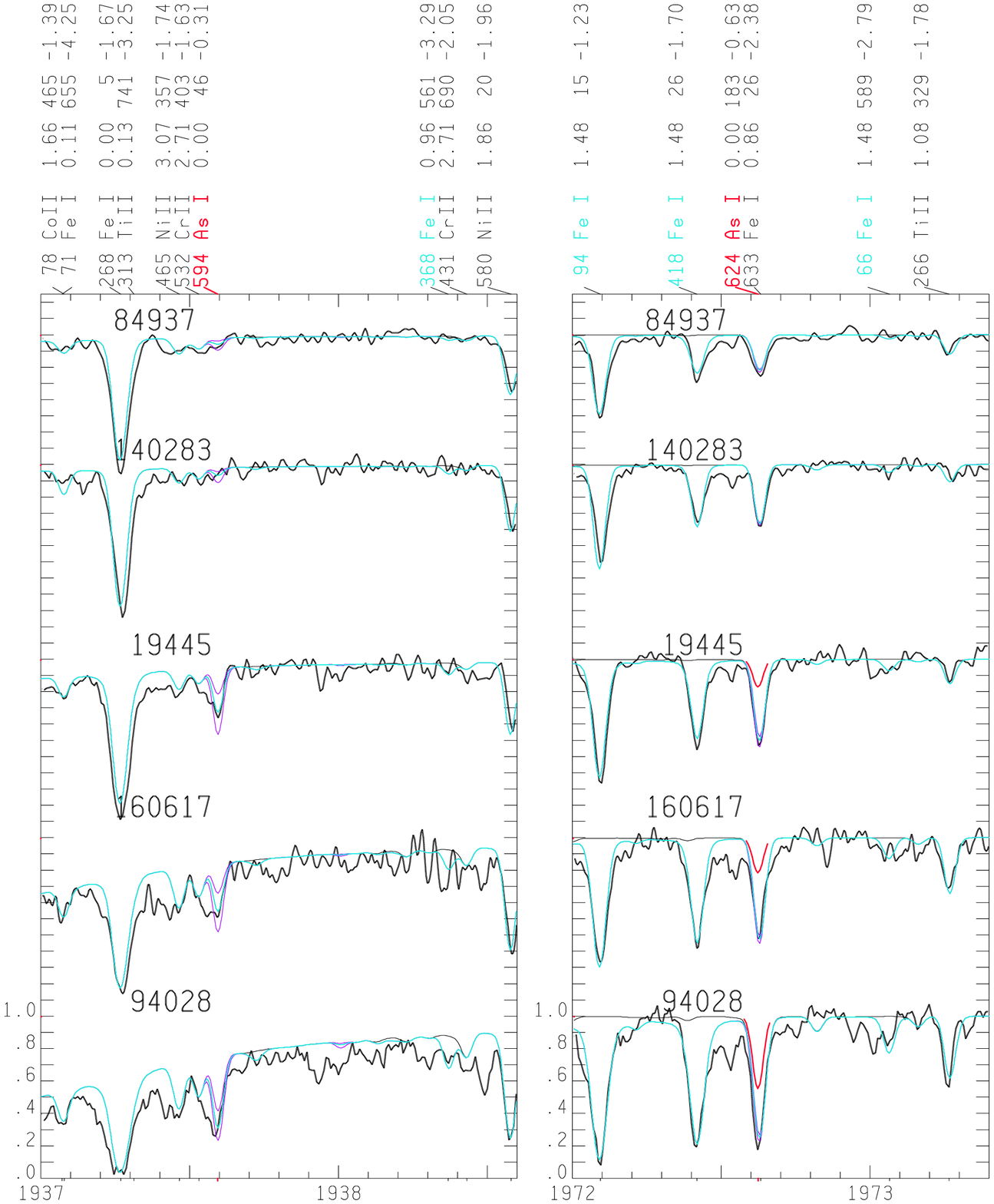}}
%\centering
%\resizebox{\hsize}{!}{\includegraphics{figb193197.eps}}
\caption{%[]{
Spectra 
near the \ion{As}{I} lines at 1937.594 {\rm \AA} 
(left panel) 
and at 1972.624 {\rm \AA} (right panel).
} 
\label {fig3}
\end{figure}

\begin{figure}%[p]
%\resizebox{10.0cm}{12.0cm}
{\includegraphics [scale=0.40]{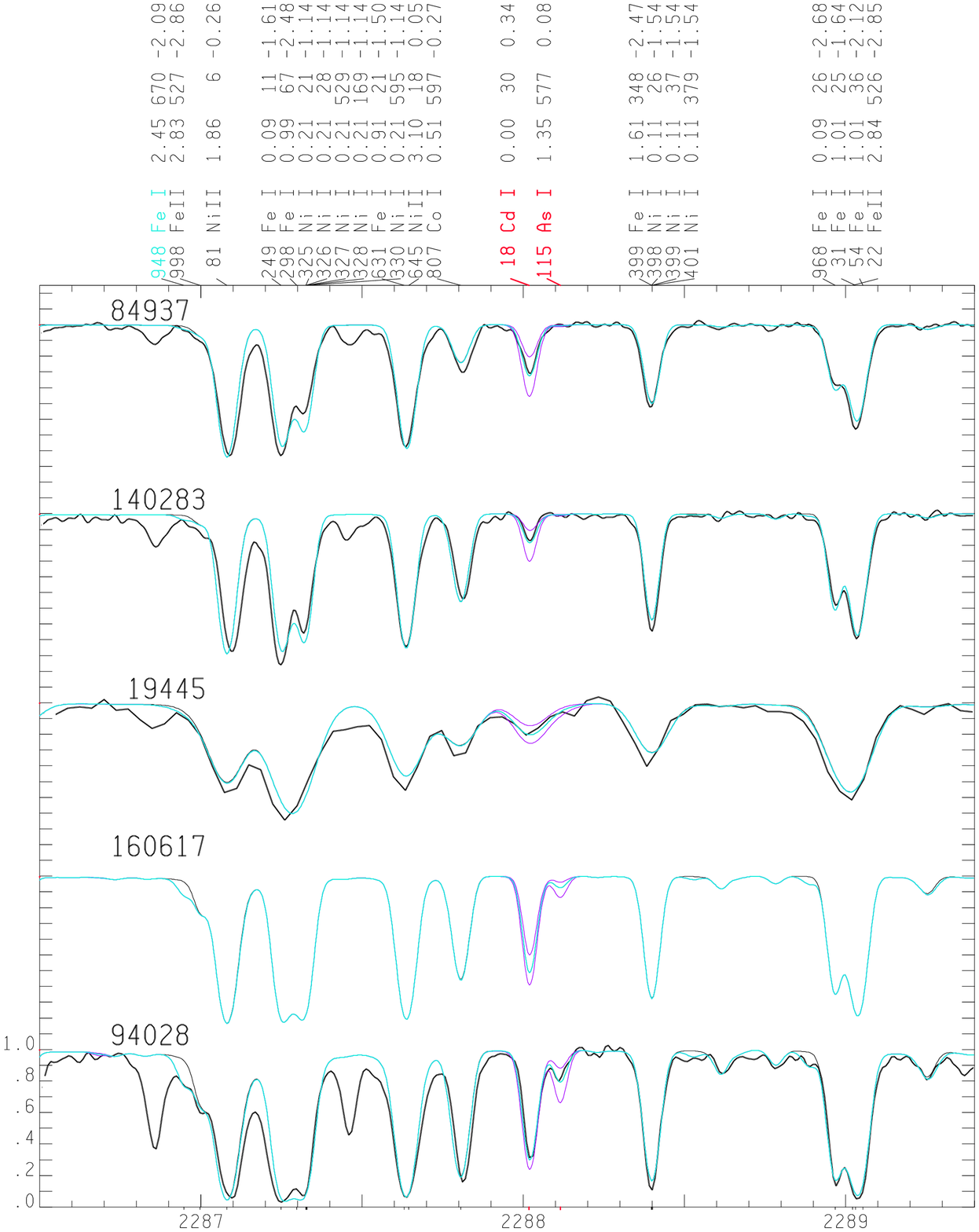}}
%\resizebox{\hsize}{!}{\includegraphics{figb228.eps}}
\caption{%[]{
Spectra near the \ion{Cd}{I} 2288.018 {\rm \AA}  line; that of HD~19445 is E230M. Data for HD 160617 are lacking in this region.}
\label {fig4}
\end{figure}

\begin{figure}
%\resizebox{8.0cm}{12.0cm}
{\includegraphics [scale=0.40]{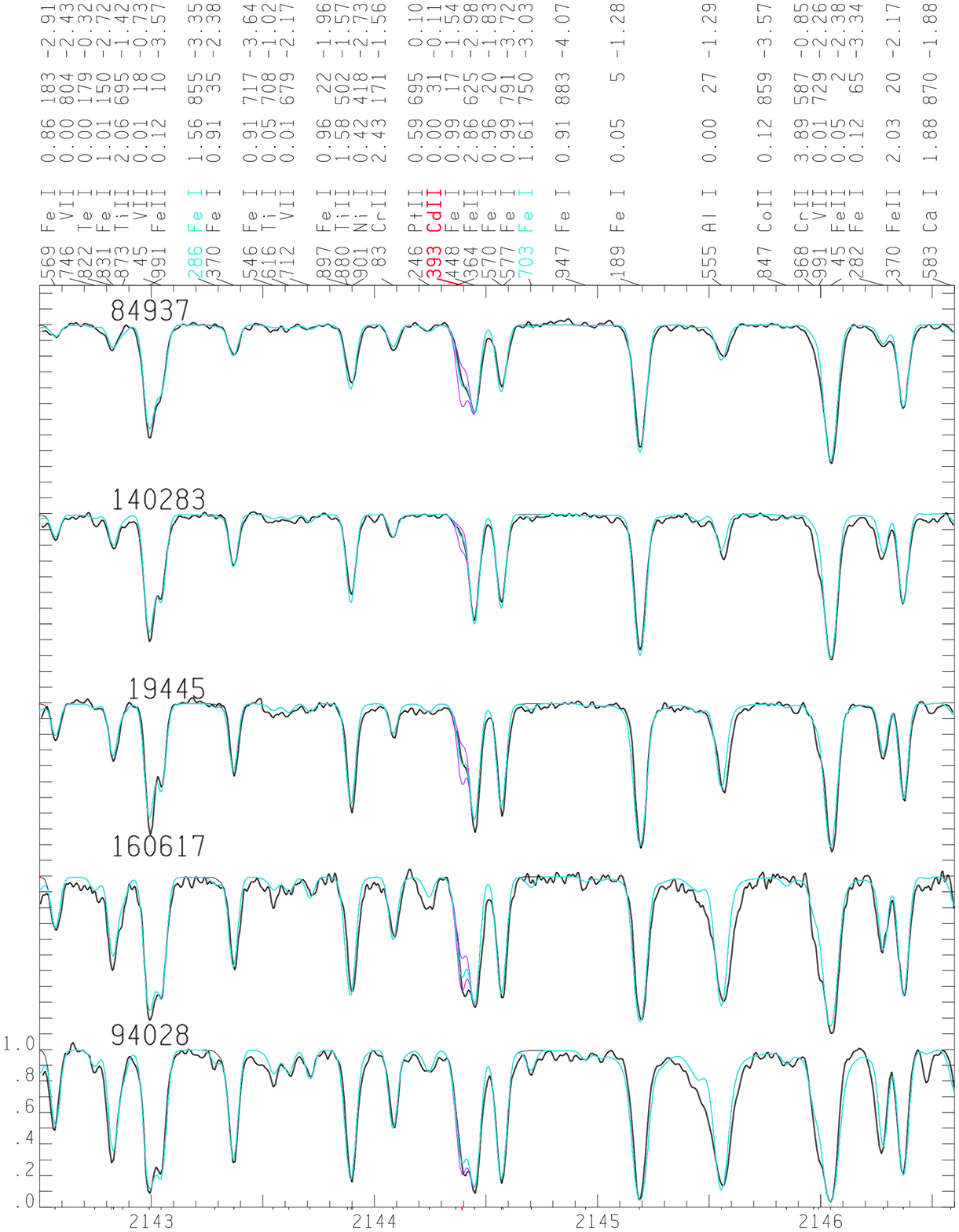}}
%\resizebox{\hsize}{!}{\includegraphics{figb214.eps}}
\caption{Spectra in the region of the \ion{Cd}{II} 2144.393  {\rm \AA} line.
The poor fit for the two bottom spectra illustrates the effect of omitting its hyperfine splitting for this plot.}
\label {fig5}
\end{figure}

Figures \ref{fig1} to \ref{fig15} 
illustrate the agreement between the observed 
spctrum and two {\tt SYNTHE} 
spectral calculations for each of the five stars 
in UV regions containing lines of light trans-Fe 
elements. These five comparisons are offset 
vertically for clarity. 

The goodness of fit of the abundance deduced for each star 
can be judged
by how well the light blue line 
representing the best-fit theoretical calculation that includes 
the newly identified \ion{Fe}{I} lines below 
matches the dark black observation. 
In the electronic version of the paper the figures can be enlarged and allow a better
appreciation of the quality of the fit.

Labels highlighted in light blue indicate %} 
lines of {\ion{Fe}{I}} newly identified in ongoing work 
by \citet{PetersonK15} and \citet{PetersonKA17}, 
including results for forty more levels since identified by \citet{PetersonKA20}.
The effects of these recently identified \ion{Fe}{I} lines are discerned
wherever the light blue line falls below the thin black line.
Lines that remain unidentified are evident wherever the solid black line
falls below the blue line in all stars, and increase in depth from the
top to the bottom spectrum. 

Fig.\  \ref{fig1} shows examples of both
types. Three newly identified lines appear at 1898.284 {\rm \AA},
1898.727 {\rm \AA}, and 1899.822 {\rm \AA}.
Two of the strongest lines that remain unidentified
are seen at 1897.78 {\rm \AA} and 1900.02 {\rm \AA}. 
Many weaker unidentified lines are also present, to a degree 
that the continuum is no longer discernible in the bottom spectrum.
Figs.\ \ref{fig2} and \ref{fig3} also show the presence of 
unidentified lines, although not quite to the same degree. 
Fig.\ \ref{fig3} contains two strong and two weaker 
newly identified {\ion{Fe}{I}} lines, as identified at the top.

%TABLE 3
\begin{table*}
\scalefont{0.68}
\caption{Lines and line parameters adopted are given, along with abundances A(X) derived from them %(with (A(X)= log (N(X)/N(H)) + 12) 
 for each star analyzed in this work. }
% The set of A(X) values listed on the left were derived with {\tt Turbospec}; those to the right were derived with {\tt SYNTHE}.
% Both analyses adopted the line parameters appearing in the first five columns. 
%Wavelengths are in vacuum for $\lambda < 2000\AA$ and in air for $\lambda > 2000\AA$.
%References for the gf values indicated in column 5: 1: \citet{LiNPW99};~~2: \citet{Morton00};~~3: \citet{Peterson11}; ~~4: \citet{HLGBW82};~~5: \citet{LNAJ06}; 6: \citet{Moehring06};~~7: \citet{SnedenCL03}. }
\label{linelist}
\centering
\begin{tabular}{lcccccccccccccccc}
%\begin{tabular}{l@{~}c@{~}c@{~~}r@{~~}c@{~~}c@{~~~~~}c@{~~~}c@{~~~}c@{~~~}c@{~~}c@{~~}c@{~~~~~}c@{~~~}c@{~~~}c@{~~~}c@{~~}c@{~~}r@{}r@{}r@{}}
%}
%\hline
%\noalign{\smallskip}
&\multicolumn{4}{c}{Line Parameters}&  &\multicolumn{5}{c}{A(X) {\tt Turbospectrum}}& &\multicolumn{5}{c}{A(X) {\tt SYNTHE}}\\
%%&\multicolumn{4}{c}{Line Parameters}&~~~~~&\multicolumn{5}{c}{A(X) {\tt Turbospectrum}}&  &~~~~~~~~~~~~~~~~~~{A(X) {\tt SYNTHE}}\\
Species & $\lambda$ & \hbox{$\chi_{ex}$(eV)} &\loggf& Ref& &19445&84937&94028&140283&160617& &19445&84937&94028&140283&160617\\
\hline
\noalign{\smallskip}
%Ge  I & 1998.887 & 0.17 & -0.435 & 1 &  & {\bf 1.00}  & {\bf 0.88}    & {\bf 2.03}   &     &{\bf 1.13}       &  & 1.16 & {\bf 0.93}       & {\bf 1.95} &{\bf$<$-0.16}  & 1.16\\
%Ge  I & 2041.712 & 0.00 & -0.54  & 1 &  & 1.00        &  0.78          & 1.83        &     & 1.13            &  & 1.04 & {\bf 0.82}       & 1.69 & {\bf 0.07} & 1.192\\
%Ge  I & 2065.215 & 0.07 & -0.84  & 1 &  & {\bf 1.23}  & {\bf $<$0.88}  & 2.03         &      & {\bf $<$1.13} &  & {\bf 1.12} & {\bf 1.00} & 1.97 & {\bf $<$-0.19} & 1.123\\
%Ge  I & 2094.258 & 0.17 & -0.12  & 1 &  & {\bf $<$1.00} & {\bf $<$0.88}  & {\bf 1.53} &      & {\bf $<$1.13} &  & 0.92 & {\bf 0.85}       & 1.72 & {\bf -0.39} & 1.020\\
%Ge  I & 2592.534 & 0.07 & -0.525 & 1 &  &  &  &  &  &  &                                                        & 1.16 & {\bf 0.84}       & 1.86 &{\bf $<$-0.15}  &  \\
%Ge  I & 2651.568 & 0.00 & -0.68  & 1 &  &  &  &  &  &  &                                                        & {\bf 1.15} & {\bf 0.93} & 1.90 & {\bf $<$-0.06} &  \\
%Ge  I & 2709.624 & 0.07 & -0.64  & 1 &  &  &  &  &  &  &                                                  & {\bf $<$1.27} &{\bf  $<$0.95} & 1.92 &{\bf $<$-0.04}  &  \\
%Ge  I & 3039.067 & 0.88 & -0.08  & 1 &  & 1.06        & 1.13           & 1.92        &-0.35  &     &            & 1.25 &{\bf 0.93}        & 1.85 &{\bf -0.16} &  \\
Ge  I & 1998.887 & 0.17 & -0.435 & 1 &  & 1.00  & 0.88    &  2.03   &     & 1.13       &  & 1.16 &  0.93       &  1.95 & $<$-0.16  & 1.16\\
Ge  I & 2041.712 & 0.00 & -0.54  & 1 &  & 1.00        &  0.78          & 1.83        &     & 1.13            &  & 1.04 &  0.82       & 1.69 &  0.07 & 1.192\\
Ge  I & 2065.215 & 0.07 & -0.84  & 1 &  &  1.23  &  $<$0.88  & 2.03         &      &  $<$1.13 &  &  1.12 &  1.00 & 1.97 &  $<$-0.19 & 1.123\\
Ge  I & 2094.258 & 0.17 & -0.12  & 1 &  & $<$1.00 &  $<$0.88  &  1.53 &      &  $<$1.13 &  & 0.92 &  0.85       & 1.72 &  -0.39 & 1.020\\
Ge  I & 2592.534 & 0.07 & -0.525 & 1 &  &  &  &  &  &  &                                                        & 1.16 &  0.84  & 1.86 & $<$-0.15  &  \\
Ge  I & 2651.568 & 0.00 & -0.68  & 1 &  &  &  &  &  &  &                                                        &  1.15 &  0.93 & 1.90 &  $<$-0.06 &  \\
Ge  I & 2709.624 & 0.07 & -0.64  & 1 &  &  &  &  &  &  &                                                  &  $<$1.27 &  $ <$0.95 & 1.92 & $<$-0.04  &  \\
Ge  I & 3039.067 & 0.88 & -0.08  & 1 &  & 1.06   & 1.13           & 1.92        &-0.35  &     &            & 1.25 & 0.93   & 1.85 & -0.16 &  \\
%\hline
\bf{Mean A(Ge)} &  &  &  &  & ~~ & \bf{1.06} & \bf{$<$0.91}       & \bf{1.87} & \bf{-0.35} & \bf{$<$1.15}   &  & \bf{1.13} & \bf{0.91}   & \bf{1.86} & \bf{$\leq$-0.15} & \bf{1.12}\\
StDev &  &  &  &  &  & 0.13 & 0.08 & 0.15 &     &     &                                                         & 0.12 & 0.07             & 0.10 & 0.14 & 0.07\\
%No. Lines &  &  &  &  &  &  &  &  &  &  &  & 7 & 7 & 8 & 4 & 4\\
\\
%\hline
As  I & 1890.429 & 0.000 & -0.068 & 2 &  & 0.15 & -0.65 & 1.20 & -0.90 & 0.00 &  & 0.31 & -0.74 & 1.01 & -0.99 & 0.11\\
As  I & 1937.594 & 0.000 & -0.307 & 2 &  & 0.38 & -0.45 & 1.20 & -0.90 & 0.50 &  & 0.31 & -0.54 & 1.01 & -0.99 & 0.21\\
%\hline
\bf{Mean A(As)} &  &  &  &  & ~~ & \bf{0.27} & \bf{-0.51}& \bf{1.20} & \bf{-0.80} & \bf{0.25} &             & \bf{0.31} & \bf{-0.64}   & \bf{1.01} & \bf{-0.99} & \bf{0.16}\\
StDev &  &  &  &  &  & 0.16 & 0.09 & 0.000 & 0.14 & 0.35 &  & 0.00 & 0.14 & 0.00 & 0.00 & 0.07\\
%No. Lines &  &  &  &  &  & 2 & 2 & 2 & 2 & 2 &  & 2 & 2 & 2 & 2 & 2\\
\\
%\hline
Se  I & 1960.893 & 0.000 & -0.434 & 2 &  & 1.19 & 0.89 & 2.24 & 0.54 & 1.54 &  & 1.65 & 1.20 & 2.35 & 0.45 & 1.65\\
Se  I & 2039.842 & 0.25 & -0.737 & 2 &  & 1.49 & 1.29 & 2.24 & 0.64 & 1.54 &  & 1.75 & 1.30 & 2.35 & 0.65 & 1.85\\
Se  I & 2074.794 & 0.000 & -2.261 & 2 &  & 1.49 & 1.29 & 2.24 & 0.84 & 1.54 &  & 1.75 & 1.40 & 2.35 & 0.75 & 1.80\\
%\hline
\bf{Mean A(Se)} &  &  &  &  & ~~ & \bf{1.39} & \bf{1.16} & \bf{2.24} & \bf{0.67} & \bf{1.54} &  & \bf{1.72} & \bf{1.30} & \bf{2.35} & \bf{0.62} & \bf{1.77}\\
StDev &  &  &  &  &  & 0.17 & 0.23 & 0.00 & 0.15 & 0.00 &  & 0.06 & 0.10 & 0.00 & 0.15 & 0.10\\
%No. Lines &  &  &  &  &  & 3 & 3 & 3 & 3 & 3 &  & 3 & 3 & 3 & 3 & 3\\
\\
%\hline
Sr II & 4077.709 & 0.000 & 0.147 & 2 &  & 1.02 & 1.08 & 1.82 & 0.12 & 1.31 &  & 1.13 & 0.88 & 1.73 & 0.08\\
Sr II & 4215.519 & 0.000 &-0.176 & 2 &  & 0.99 & 1.01 & 1.85 & 0.03 & 1.57 &  & 1.14 & 0.89 & 1.74 & 0.09\\
%\hline
\bf{Mean A(Sr)} &  &  &  &  & ~~ & \bf{1.00} & \bf{1.04} & \bf{1.83} & \bf{0.07} & \bf{1.36} &  & \bf{1.14} & \bf{0.89} & \bf{1.74} & \bf{0.09}\\
StDev &  &  &  &  &  &                 0.03     & 0.05      & 0.02      & 0.07      & 0.07 &  & 0.01     & 0.01      & 0.01      & 0.01 & \\
%No. Lines &  &  &  &  &  & 2 & 2 & 2 & 2 & 2 &  & 2 & 2 & 2 & 2\\
\\
%\hline
Y II & 3600.732 & 0.18  &  0.28 & 3 &  & 0.10 & -0.03 & 0.87 & -0.78 & 0.32 &  & 0.18 & 0.03 & 0.98 & -0.82 & 0.43\\
Y II & 3601.916 & 0.104 & -0.18 & 3 &  & 0.10 & -0.03 & 0.87 & -0.81 & 0.32 &  & 0.23 & 0.13 & 0.98 & -0.87 & 0.38\\
Y II & 3611.043 & 0.13  &  0.11 & 3 &  & 0.07 & -0.05 & 0.87 & -0.89 & 0.30 &  & 0.18 & -0.07& 0.98 & -0.87 & 0.28\\
Y II & 3788.694 & 0.104 & -0.07 & 3 &  & 0.14 & -0.03 & 0.94 & -0.74 & 0.44 &  & 0.28 & 0.08 & 1.03 & -0.82 & 0.48\\
Y II & 3818.341 & 0.13  & -0.98 & 3 &  &      &       & 0.92 &       & 0.32 &  &      &      & 0.98 &       &   \\
Y II & 3950.349 & 0.104 & -0.49 & 3 &  & 0.20 & -0.10 & 0.91 & -0.67 & 0.42 &  & 0.28 & 0.08 & 0.98 & -0.82 &   \\
%\hline
\bf{Mean A(Y)} &  &  &  &  & ~~ & \bf{0.12} & \bf{-0.05}& \bf{0.90} & \bf{-0.78} & \bf{0.35} &  & \bf{0.23} & \bf{0.05} & \bf{0.99} & \bf{-0.84} & \bf{0.39}\\
StDev &  &  &  &  &                & 0.05      & 0.03      & 0.03      & 0.08       & 0.06      &  & 0.05      & 0.08      & 0.02      & 0.03       & 0.09\\
%No. Lines &  &  &  &  &  & 5 & 5 & 6 & 5 & 6 &  & 5 & 5 & 6 & 5 & 4\\
\\
%\hline
Zr II & 3438.231 & 0.095 & 0.41  & 4 &  & 0.76 & 0.65 &      & -0.11 & 1.02 &  & 0.89 & 0.74 & 1.59 & -0.16 & 1.09\\
Zr II & 3496.205 & 0.039 & 0.26  & 4 &  & 0.79 & 0.67 & 1.43 & -0.12 & 1.04 &  & 0.92 & 0.72 & 1.57 & -0.18 & 1.07\\
Zr II & 3551.951 & 0.095 & -0.36 & 4 &  & 0.85 & 0.69 & 1.61 & -0.07 & 1.07 &  & 0.94 & 0.79 & 1.74 & -0.06 & 1.19\\
Zr II & 3556.594 & 0.466 & 0.07  & 4 &  & 0.74 & 0.57 & 1.45 & -0.19 & 0.96 &  & 0.91 & 0.76 & 1.66 & -0.14 & 1.16\\
Zr II & 3576.853 & 0.409 & -0.12 & 4 &  & 0.88 & 0.70 &      & -0.02 & 1.13 &  & 0.98 & 0.83 & 1.63 & 0.03  & 1.18\\
Zr II & 3614.765 &       & -0.25 & 4 &  & 0.81 & 0.64 & 1.63 &       & 1.05 &  &      &      &      &       &   \\
Zr II & 3836.761 & 0.559 & -0.12 & 4 &  & 0.77 & 0.61 &      & -0.09 & 1.00 &  & 0.94 & 0.69 & 1.64 & -0.16 & 1.14\\
Zr II & 4149.198 & 0.802 & -0.04 & 4 &  & 0.85 & 0.66 & 1.61 & -0.12 & 1.12 &  &      &      &      &       &   \\
Zr II & 4161.200 & 0.713 & -0.59 & 4 &  & 0.91 &      & 1.57 & -0.08 & 1.13 &  &      &      &      &       &   \\
Zr II & 4208.980 & 0.713 & -0.51 & 4 &  & 0.87 &      & 1.60 & -0.02 & 1.13 &  &      &      &      &       &   \\
%\hline
{\bf Mean A(Zr)} &  &  &  &  & ~~ & \bf{0.82} & \bf{0.65} & \bf{1.56} & \bf{-0.05}& \bf{1.07} &  & \bf{0.93} & \bf{0.76} & \bf{1.64} & \bf{-0.11}& \bf{1.14}\\
StDev &  &  &  &  &                  & 0.06      & 0.04      & 0.08      & 0.10      & 0.06      &  & 0.03      & 0.05      & 0.06      & 0.08      & 0.05\\
%No. Lines &  &  &  &  &  & 6 & 6 & 6 & 6 & 6 &  & 6 & 6 & 6 & 6 & 6\\
\\
%\hline
Mo II & 2015.123 & 0.000 & -0.362 & 5 &  & 0.45 & 0.19 & 1.47 & -0.50 & 0.66 &  & 0.56 & 0.31 & 1.46 & -0.64 & 0.76\\
Mo II & 2020.324 & 0.000 &  0.155 & 5 &  & 0.51 & 0.23 & 1.29 & -0.60 & 0.57 &  & 0.56 & 0.31 & 1.46 & -0.64 & 0.76\\
Mo II & 2038.452 & 0.000 & -0.205 & 5 &  & 0.45 & 0.28 & 1.20 & -0.68 & 0.53 &  & 0.56 & 0.31 & 1.46 & -0.64 & 0.76\\
Mo II & 2045.984 & 0.000 & -0.22  & 5 &  & 0.47 & 0.28 & 1.30 & -0.53 & 0.57 &  & 0.56 & 0.31 & 1.46 & -0.64 & 0.76\\
Mo II & 2081.696 & 0.000 & -1.15  & 5 &  &      &      &      &       &      &  & 0.66 & 0.31 & 1.56 & -0.54 & 0.81\\
Mo  I & 3864.104 & 0.000 & -0.01  & 5 &  & 0.49 &      & 1.02 &       & 0.60 &  &      &      & 1.26 &       & 0.86\\
%\hline
\bf{Mean A(Mo)} &  &  &  &    & ~~ & \bf{0.45} & \bf{0.25} & \bf{1.26} & \bf{-0.58} & \bf{0.59} &  & \bf{0.58} & \bf{0.31} & \bf{1.44} & \bf{-0.62} & \bf{0.79}\\
StDev &  &  &  &  &                      & 0.07      & 0.04      & 0.16      & 0.08       & 0.05 &  & 0.05      & 0.00      & 0.10      & 0.05       & 0.04\\
%No. Lines &  &  &  &  &  & 5 & 4 & 5 & 4 & 5 &  & 5 & 5 & 6 & 5 & 6\\
\\
Ru II& 1888.064 &  0.000&-0.534 & 1   &  &  0.38&  0.53& 0.74& $<$-0.30&  0.45 &  &  $<$0.18&  0.43& 1.08& $<$-0.62&  0.58  \\
Ru I & 3498.942 &  0.000& 0.310 & 1   &  &  0.50&  0.40& 1.03& $<$-0.54&  0.65 &  &  &  & & &       \\
%\hline
\bf{Mean A(Ru)} &     &     &     &    & ~~~~~ &\bf{0.44}& \bf{0.47}& \bf{0.89} & \bf{$<$-0.42} &\bf{0.55} &  &\bf{$<$}0.18& \bf{0.43}& \bf{1.08} & \bf{$<$-0.62} &\bf{0.58}\\
StDev           &     &     &     &    &       & 0.08 &  0.09& 0.21  &  0.17 & 0.14 &       &  &  &  &   &   \\
%%No. Lines &  &  &  &  &  & 2 & 2 & 2 & 2 & 2 &  &\\
\\
%\hline
Cd II& 2144.393 &  0.000& 0.018 & 6  &  & -0.49& -0.43& 0.31& -1.44& 0.05&  & -0.27 & -0.37 & 0.38 & -1.52 & 0.38\\
Cd I & 2288.018 &  0.000& 0.336 & 7   &  & -0.49& -0.39& 0.27& -1.48&    &  & -0.19 & -0.24 & 0.41 & -1.39 &     \\
%\hline
\bf{Mean A(Cd)} &  &   &    &   & ~~ &{\bf -0.49}& {\bf -0.41}& {\bf 0.29} & {\bf -1.46} &{\bf 0.05} &  & \bf{-0.23} & \bf{-0.30} & \bf{0.40} & \bf{-1.45} & \bf{0.38}\\
StDev      &     &     &     &    &       & 0.00 &  0.03& 0.03  &  0.03 &     &  & 0.05      & 0.09      & 0.02      & 0.09       &     \\
\\
%\hline
Sn II& 1899.898 &  0.527 & -0.195& 2 & ~~~~~& 0.09& -0.46& 1.14& -1.06& 0.22 & & 0.04& $<$-0.21& 1.04& $<$-1.06& 0.24 \\
\bf{A(Sn)} &     &     &     &    & ~~~~~ &\bf{0.09}& \bf{-0.46}& \bf{1.14} & \bf{-1.06} &\bf{0.22} &  &\bf{0.04}& \bf{$<$-0.21}& \bf{1.04} & \bf{$<$-1.06} &\bf{0.24}\\
\\
%\hline
Ba II & 4554.033 & 0.000 &  0.156 & 2 &  &  0.03 & -0.22 & 0.99 & -1.19 & 0.72 &  & 0.17 & -0.06 & 0.93 & -1.31 & 0.77\\
Ba II & 4934.077 & 0.000 & -0.159 & 2 &  & -0.01 & -0.25 & 0.87 & -1.22 & 0.65 &  & 0.07 & -0.16 & 0.83 & -1.31 & 0.67\\
Ba II & 5853.670 & 0.604 & -1.01  & 8 &  &       &       & 0.99 &       & 0.57 &  &  &  &  &  & \\
Ba II & 6141.713 & 0.704 & -0.07  & 8 &  & -0.01 & -0.20 & 1.10 & -0.99 & 0.70 &  &  &  &  &  & \\
Ba II & 6496.900 & 0.604 & -0.407 & 8 &  &       &       & 1.14 & -0.96 & 0.52 &  &  &  &  &  & \\
%\hline
\bf{Mean A(Ba)} &  &  &  &  & ~~ & \bf{0.0 }  & \bf{-0.22} & \bf{1.02} & \bf{-1.09} & \bf{0.63} &  & \bf{0.12}  & \bf{-0.11} & \bf{0.88} & \bf{-1.31} & \bf{0.72}\\
StDev &  &  &  &  &                 & 0.03       & 0.03       & 0.11      & 0.13       & 0.09    &    & 0.07       & 0.07       & 0.07      & 0.00       & 0.07\\
%No. Lines &  &  &  &  &  & 3 & 3 & 5 & 4 & 5 &  & 2 & 2 & 2 & 2 & 2\\
\\
%\hline
Eu II & 3930.4   & 0.000 & 0.270 & 8 & ~~~~~ &        &        &        &        & -0.77\\
Eu II & 3971.9   & 0.210 & 0.270 & 8 & ~~~~~ &        &        & -0.75  &        & -0.76\\
Eu II & 4129.7   & 0.210 & 0.220 & 8 & ~~~~~ & -1.26  & -1.35  & -0.72  & -2.27  & -0.77\\
\bf{Mean A(Eu)} & &      &       &   & ~~~~~ & \bf{-1.26}  & \bf{-1.35} & \bf{-0.73}  & \bf{-2.27}  & \bf{-0.77}\\
StDev      &     &       &       &   & ~~~~~ &             &            & 0.02        &             & 0.01\\
\hline
% Use \tablefoot and \tablebib
%\noalign{\smallskip}
%\multicolumn{17}{l} {Notes: A(X)= log (N(X)/N(H)) + 12. A(X) values on the left were derived with {\tt Turbospec}; those on the right, with {\tt SYNTHE}.} \\
%\multicolumn{17}{l} {Blank entries for A(X) indicate lines lacking data or considered too unreliable for inclusion.}\\
%\multicolumn{15}{l} {~~~~those on the right, with {\tt SYNTHE}.} \\
%\multicolumn{15}{l} {$\rm [Fe/H]=  log(N_{Fe} / N_{H})_{star} - log(N_{Fe} / N_{H})_{Sun}$} \\
%\multicolumn{17}{l} {Wavelengths are in vacuum for $\lambda < 2000\AA$ and in air for $\lambda > 2000\AA$. $\rm [Fe/H] =  log(N_{Fe} / N_{H})_{star} - log(N_{Fe} / N_{H})_{Sun}$}\\
%\multicolumn{17}{l} {References for gf values (col.\ 5): 1: \citet{LiNPW99};~~2: \citet{Morton00};~~3: \citet{HLGBW82};~~4: \citet{LNAJ06};~~5: \citet{Peterson11};}\\
%\multicolumn{17}{l} {~~~~~6: \citet{RoedererLawler12};~~7: \citet{Warner68b}; ~~8: \citet{SnedenCL03}}\\
%\multicolumn{15}{l} {~~~~4: \citet{HLGBW82};~~5: \citet{LNAJ06}; 6: \citet{Moehring06};~~7: \citet{SnedenCL03}} \\
\hline
\end{tabular}
\tablefoot{A(X)= log[(N(X)/N(H)] + 12. A(X) values on the left were derived with {\tt Turbospec}; those on the right, with {\tt SYNTHE}. %\\
%{\bf A(X)= log[(N(X)/N(H)] + 12. A(X) values on the left were derived with {\tt Turbospec}; those on the right, with {\tt SYNTHE}. %\\
%Blank entries for A(X) indicate lines lacking data or considered too unreliable for inclusion. %\\
$\rm [Fe/H]=  log(N_{Fe} / N_{H})_{star} - log(N_{Fe} / N_{H})_{Sun}$. %\\
Wavelengths are in vacuum for $\lambda < 2000\AA$ and in air for $\lambda > 2000\AA$.} %\\
%}
\tablebib{(1)~\citet{LiNPW99}; (2) \citet{Morton00}; (3) \citet{HLGBW82}; (4) \citet{LNAJ06}; 
%{\bf (1)~\citet{LiNPW99}; (2) \citet{Morton00}; (3) \citet{HLGBW82}; (4) \citet{LNAJ06}; 
(5) \citet{Peterson11}; (6) \citet{RoedererLawler12}; (7) \citet{Warner68b}; (8) \citet{SnedenCL03}}
%}
\end{table*}
Straightforward theoretical considerations suggest that 
these three regions blueward of 2000\AA\
should harbor considerably more unidentified {\ion{Fe}{I}} lines
than do the regions somewhat redward of 2000\AA, as follows.
According to \citet{PetersonK15}, 
in the near UV, most unidentified lines in cool stars are expected to be due to {\ion{Fe}{I}}. 
They are transitions from a lower level that has been identified, to an upper level 
that is unidentified, since {\ion{Fe}{I}} levels below $\approx$ 7 eV have largely been 
identified. 
The ground state of a species is defined as 0 eV from its lowest level; that of {\ion{Fe}{I}} is 
of even parity. Lines of {\ion{Fe}{I}} have the upper level of the opposite parity as the 
lower level. Consequently, the number of unidentified lines remaining whose upper level 
is odd increases dramatically only below $\approx$ 2000\AA, as seen in Fig.\ \ref{fig1}. 

A similar increase in the number of unidentified lines is expected redward of 2500\AA. 
This is because the lowest {\ion{Fe}{I}} odd-parity 
levels are not at zero but at several eV. The strong presence of unidentified 
lines in the 2630 -- 2720\AA\ region where weak lines prevail 
is indeed observed from the low-resolution UV spectra of dwarfs. 
Throughout this region, unmodeled flux is absent only in the most 
metal-poor dwarfs of near-solar temperature, and climbs to dramatic 
levels at solar abundances \citep[Fig.\ 4]{PetersonK15}.

That these are largely due to {\ion{Fe}{I}} is confirmed by 
the Kurucz "predicted" calculations of unidentified 
{\ion{Fe}{I}} lines, those for which the energy of one level (or both) is unknown.
Currently, the Kurucz predicted {\ion{Fe}{I}} calculations show more than fifty unidentified
{\ion{Fe}{I}} lines remaining in the 2650 -- 2700\AA\ region
with predicted strengths greater than 20\% deep in the HD~94028 spectrum
of Fig.\ \ref{fig11}, and roughly five times that number greater than 10\% deep. 
Because of their higher low excitation potentials, their strengths are considerably 
reduced (by the exponential Boltzmann excitation factor) compared to those $<$2000\AA. 
So they are much more difficult to identify, especially in the laboratory 
\citep{BrownGJ88}, but their numbers remain strong.

The effect of these unidentified lines on abundances derived from weak 
lines in this region is thus expected to be large, on both the contamination of weak 
features and on the placement of the local continuum. It is best judged 
from Fig.\ \ref{fig11}, where many weak newly identified and unidentified 
lines are both present. Within this 3\AA\ range, five newly identified 
{\ion{Fe}{I}} lines are strong enough to be labeled at the top, and weaker ones such 
as those at 2650.66\AA\ and 2652.01\AA\ are seen in the HD\,94028 spectrum. 

Blending by newly identified {\ion{Fe}{I}} lines does impact abundance determinations 
for trans-Fe elements. One such case is seen in Fig.\ \ref{fig11}. 
Two {\ion{Fe}{I}} lines \citet{PetersonKA17} identified at 2651.102\AA\ and 
2651.182\AA\ blend with one of the strongest {\ion{Ge}{I}} lines,
at 2651.172\AA. The extent of their contamination is judged 
by the difference in the depth of the blue line, which includes 
them in the synthesis, versus that of the thin black line, which does not. 
This is seen most clearly 
in the plots for HD\,140283 and HD\,160617.
In HD\,160617, 
the black line falls higher than 
does the weaker of the two red lines, indicating that the analysis of 
this line would yield a germanium abundance about 0.4 dex too high 
if the {\ion{Fe}{I}} contribution was ignored. While the {\tt SYNTHE} calculations 
do include this contamination, and produce consistent results for germanium 
for the other stars, we dropped this line due to its blending. 
Its potentially unrecognized contamination highlights the need for 
including as many well-modeled lines as possible for species reliant 
upon UV spectra.

Another case is seen in Fig.\ \ref{fig9}, even though unidentified 
lines are much less common. In contrast to Fig.\ \ref{fig11}, 
the 5\AA\ region in Fig.\ \ref{fig9} has but a single 
newly identified {\ion{Fe}{I}} line, at 2062.800\AA. However, as it lies close to the 
{\ion{Se}{I}} line at 2062.788\AA, its contamination is also significant, 
and so this line was also dropped. 

Those lines 
in Fig.\ \ref{fig11} that remain unidentified are mostly 
too weak to affect continuum placement in the E230H spectrum
of the warm star HD~84937 at [Fe/H] = $-$2.25. 
Unidentified lines do become significant in the continuum placement
of the lower resolution E230M spectrum of the cooler star HD~19445
at [Fe/H] = --2.0, and also in the E230H spectrum of HD\,94028
at [Fe/H] = --1.4, as seen in the increasing number and strength of
black depressions that have no counterparts in the blue calculation. 
In Fig.\ \ref{fig9}, weaker depressions are seen in HD 94028; these are 
only barely discerned in HD 160617 and they are absent from the other stars. 

In lower-resolution E230M spectra of cool metal-poor giants, 
\citet{CowanSB05} noted that the UV {\ion{Ge}{I}}
lines such as those near 2651\AA\ were detectable, 
but were very difficult to analyze due to their contamination. 
Because {\ion{Fe}{I}} lines strengthen at low temperatures, it seems probable 
that their difficulties were due to the significant presence 
of unidentified {\ion{Fe}{I}} lines contaminating the surroundings.

\clearpage

\section {Abundance results}
\label{abdav}

Table \ref{axsum} presents the final A(X) values adopted for the Sun as well as for each star.
In Table \ref{axsum}, most solar values of A(X) are adopted from \citet{GrevesseSA15}; those for As and Se
are from \citet{AsplundGS09}.  
In Table \ref{totheavy}, we present the corresponding values of [X/Fe] for the {\tt Turbospec} models.

To derive A(X) for each star, the line-by-line results in {\bf boldface}  
for each element as listed in 
Table \ref{linelist} were combined as follows. 
For UV lines, each individual {\tt Turbospectrum} and {\tt SYNTHE} determination 
was averaged with equal weight. Upper limits were considered as detections, 
which could lead to erroneously high abundances in cases with many upper limits.
For lines redward of 3050\AA, the unweighted average of {\tt Turbospec} results alone was adopted, due to the higher quality of the optical spectra employed. 
Wherever lines from both {\tt Turbospec} and {\tt SYNTHE} were used, Table \ref{axsum} lists the internal standard deviation of the individual line-by-line determinations in both separate analyses. It is a measurement of the overall coherence of the two sets of computations.

Germanium warranted an exception. For germanium, we adopted the {\tt SYNTHE} calculations alone 
for the A(Ge) values in Table \ref{axsum}. The {\tt SYNTHE} calculations 
do include the newly identified {\ion{Fe}{I}} lines, whose importance for 
germanium is described at the end of Sec.\ \ref{uvfig}. In addition, the {\tt SYNTHE}
calculations over 2000\AA\ -- 2080\AA\ (Figs.\ \ref{fig7}, \ref{fig9}, \ref{fig10}, \ref{fig14}, and \ref{fig15}) 
incorporate both of the two E230H spectra available for this region in HD~140283 (Table 1). 
This aided estimates of detection for extremely weak lines, especially critical for germanium. 
As noted in Sec.\ \ref{linegfselect}, the {\tt SYNTHE} analysis yielded results whose internal 
dispersion is low, which supports the internal consistency of the \citet{LiNPW99} 
{\ion{Ge}{I}} gf values; it supports the reliability of the {\tt SYNTHE} analysis as well. However, since many 
of the {\tt SYNTHE} determinations are upper limits, 
our result for that star is best considered a possible upper limit, as indicated in Table \ref{axsum}.

\begin{table}
\caption[]{Abundances A(X) adopted for the Sun and found from Table \ref{linelist}}
\label{axsum}
\scalefont{0.95}
\centering
\begin{tabular}{l@{~}c@{~~~~~}c@{~~~}c@{~~~}c@{~~~}c@{~~}c@{~~}}
\hline
\noalign{\smallskip}
        &    &HD    & HD    & HD   & HD     &HD    \\
Element &Sun &19445 & 84937 &94028 & 140283 &160617\\
\hline
\noalign{\smallskip}
A(Ge) & 3.63 & 1.13 & 0.91 & 1.86 & $\leq$-0.15 & 1.12\\
%StDev &      & 0.04 & 0.04 & 0.09 &  0.14 & 0.07\\
\\
A(As) & 2.30 & 0.29 & -0.58 & 1.11 & -0.90 & 0.21\\
StDev &      & 0.03 &  0.09 & 0.13 &  0.13 & 0.06\\
\\
A(Se) & 3.34 & 1.56 & 1.23 & 2.30 & 0.65 & 1.66\\
StDev &      & 0.23 & 0.10 & 0.08 & 0.04 & 0.16\\
\\
A(Sr) & 2.83 & 1.07 & 0.97 & 1.79 &  0.08 & 1.36\\
%\hline
\\
A(Y)  & 2.21 & 0.12 & -0.05 & 0.90 & -0.78 & 0.35\\
%\hline
\\
A(Zr) & 2.59 & 0.82 & 0.65  & 1.56 & -0.05 & 1.07\\
\\
A(Mo) & 1.88 & 0.46 & 0.25 & 1.29 &  -0.58 & 0.59\\
StDev &      & 0.09 & 0.04 & 0.13 &  0.03 & 0.14\\
\\
%\hline
A(Ru) & 1.75 & 0.35 & 0.45 & 0.95 & {$<$}-0.49 & 0.56\\
StDev &      & 0.16 & 0.07 & 0.18 &  0.17 & 0.10\\
\\
%\hline
A(Cd) & 1.77 & -0.36 & -0.36 &  0.34 & -1.46 &  0.22\\
StDev &      &  0.15 &  0.08 &  0.06 &  0.05 &  0.23\\
\\
%\hline
A(Sn) & 2.02 &  0.07 & -0.34 &  1.09 & -1.06 &  0.23\\
StDev &      &  0.04 &  0.18 &  0.07 &  0.00 &  0.01\\
\\
%\hline
A(Ba) & 2.25 & 0.00 & -0.22 & 1.02 & -1.09 & 0.63\\
\\
%\hline
A(Eu) & 0.52 &-1.26 & -1.35 &-0.73 & -2.27 & -0.77\\
\hline
\end{tabular}
%{\bf
\tablefoot{A(X)= log[(N(X)/N(H)] + 12.}
%{\bf A(X)= log[(N(X)/N(H)] + 12.}
%}
%}
\end{table}

%TABLE 5
\begin{table}
\caption[]{Abundance ratios [X/Fe] derived from A(X) values in Table \ref{axsum}
}
\label{totheavy}
\begin{tabular}{l@{~~}c@{~~}c@{~~}c@{~~}c@{~~}c@{~~}c@{~~}c@{~~}c@{~~}c@{~~}c@{~~}}
\hline
\noalign{\smallskip}
                    & HD    &HD    & HD   & HD     & HD     \\
                & 19445 &84937   & 94028  & 140283 & 160617 \\
\hline
\noalign{\smallskip}
\hbox{[Fe/H]}   &--2.15 & --2.25 &--1.40  & --2.57 & --1.80 \\
\hline
\hbox{[Ge/Fe]}  & -0.35 &   -0.47 & -0.37  &  $\leq$-1.21 &  -0.71 \\
\hbox{[As/Fe]}  & +0.14 &  -0.63 & +0.21  &  -0.63 &  -0.29 \\
\hbox{[Se/Fe]}  & +0.37 &  +0.14 & +0.36  &  -0.12 &  +0.12 \\
\hbox{[Sr/Fe]}  & +0.39 &  +0.39 & +0.36  &  -0.18 &  +0.33 \\
\hbox{[Y/Fe]}   &  +0.06 &   -0.01 &  +0.09  &   -0.42 &  -0.06 \\
\hbox{[Zr/Fe]}  &  +0.38 &  +0.31 &  +0.37  &   -0.07 &   +0.28 \\
\hbox{[Mo/Fe]}  &  +0.73 &   +0.62 &  +0.81  &   +0.11 &   +0.51 \\
\hbox{[Ru/Fe]}  & +0.75 &  +0.95 & +0.60  &  {$<$}+0.33 &  +0.61 \\
\hbox{[Cd/Fe]}  &  +0.02 & +0.12 & -0.03  &  -0.66 &  +0.25 \\
\hbox{[Sn/Fe]}  & +0.20 &  -0.11 & +0.47  &  -0.51 &  +0.01 \\
\hbox{[Ba/Fe]}  & -0.10 &  -0.22 & +0.17  &  -0.77 &  +0.18 \\
\hbox{[Eu/Fe]}  & +0.37 &  +0.38 & +0.15  &  -0.22 &  +0.51 \\
\hline
\end{tabular}
%{\bf
\tablefoot{For element X, $\rm [X/Fe]=  log(N_{X} / N_{Fe})_{star} - log(N_{X} / N_{Fe})_{Sun}$}
%{\bf For element X, $\rm [X/Fe]=  log(N_{X} / N_{Fe})_{star} - log(N_{X} / N_{Fe})_{Sun}$}
%}
%}
\end{table}

\subsection{Comparison with previous works for these stars}
\label{compaRoed}
UV studies of most of the elements in Table \ref{totheavy}  
were conducted for  HD\,94028 by \citet{RoedererKP16} and for HD\,160617
by \citet{RoedererLawler12}.
Their results for each element generally agree with those of Table \ref{totheavy} to within the joint standard deviation of about 0.2 dex in the determinations. 

However, differences are large in 
the germanium abundance we find for the two stars in common with \citet{Roederer12}. 
For HD~140283, our Ge result A(X)= $\leq $ -0.15 in Table \ref{axsum} 
is considerably lower than the \citet{Roederer12} 
upper limit of +0.6. For HD~94028, we find a higher value for A(Ge), +1.86 
versus +1.56. \citet{Roederer12} considered only four
 {\ion{Ge}{I}} lines, obtaining upper limits for all of them in HD~140283, and one detection 
plus two upper limits in HD~94028. For the latter star, they had only a single UV 
spectrum centered at 2013\AA; for HD~140283, only an E230M spectrum for the critical 
2390 -- 3135\AA\ region.
As discussed above, our use of a more extensive  {\ion{Ge}{I}} line list with consistent gf values,
and of an expanded E230H dataset better able to detect them \citep{CowanSB05},
undoubtedly has led to more robust results for this difficult analysis.

Differences are also large for arsenic. The Roederer et al. values for [As/Fe] are higher than ours by
0.4 dex for HD\,94028 \citep{RoedererKP16} and 0.5 dex for HD\,160617 \citep{RoedererLawler12}.
As their adopted temperatures and gravities are similar to ours, 
we attribute the discrepancy to the choice of \ion{As}{I} lines, 
all of which lie blueward of 2000\AA.

We adopted only the lines at 1890.4\AA\ and 1937.6 \AA, 
plotted respectively in Fig.\ \ref{fig2} and in the left panel of Fig.\ \ref{fig3}. 
We dropped the only line adopted by \citet{RoedererKP16} for HD\,94028, as this 1990.4\AA~line is weaker and more blended. We also rejected the 1972.6\AA\ line that \citet{RoedererLawler12} included along with those at 1890.4\AA\ and 1990.4\AA\ for HD\,160617. 

The 1972.6\AA\ line was dropped because it is dominated 
by an {\ion{Fe}{I}} line that appears in the laboratory {\ion{Fe}{I}} spectrum of \citet{BrownGJ88}. 
The right panel of Fig.\ \ref{fig3} shows
{\tt SYNTHE} calculations that adopt the arsenic abundances of Table \ref{axsum}, and incorporate an {\ion{Fe}{I}} line
at the wavelength and lower excitation potential listed in Table 1 of \citet{BrownGJ88}.
The red lines in this panel depict calculations that include the {\ion{As}{I}} line only; its contribution
is too small to register in the top two spectra.
Adding the {\ion{Fe}{I}} line produces an excellent match to the strength and profile of the blended feature in each of the five stars.
Clearly the {\ion{Fe}{I}} line dominates the strength of the {\ion{As}{I}} feature in all five spectra,
and is the sole contributor in the two weakest-lined spectra.

NLTE effects due to departures from thermal
and Saha equilibrium populations may be influential among lines of
minority neutral species, including those of most trans-Fe elements. 
We are not aware of NLTE calculations for these trans-Fe neutral species, 
and would encourage theoretical calculations 
for UV resonance lines of these neutral species wherever possible. 
In this work they appear to be small.

\begin{figure}
%\resizebox{8.0cm}{12.0cm}           	
{\includegraphics [scale=0.40]{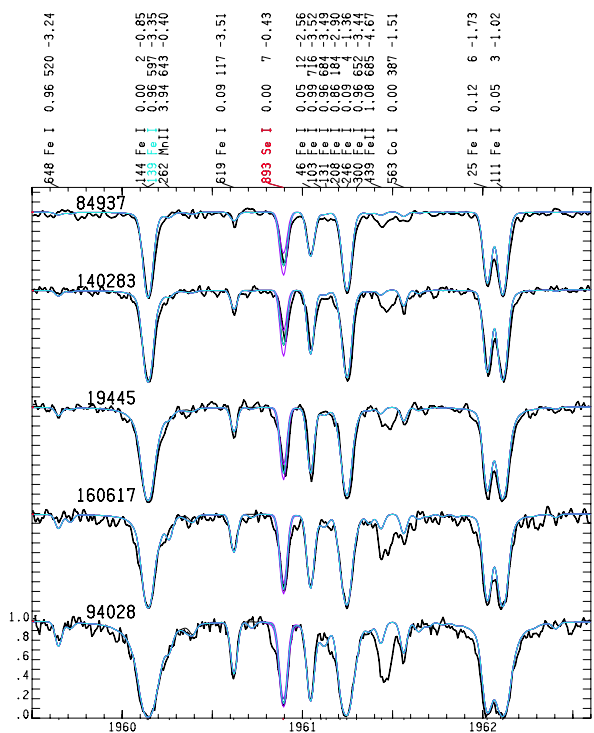}}
%\resizebox{\hsize}{!}{\includegraphics{figb195.eps}}
\caption{Spectra in the region of the \ion{Se}{I} 1960.893 {\rm \AA} line.}
\label {fig6}
\end{figure}

\begin{figure}
%\resizebox{12.0cm}{12.0cm}           	
{\includegraphics [scale=0.40]{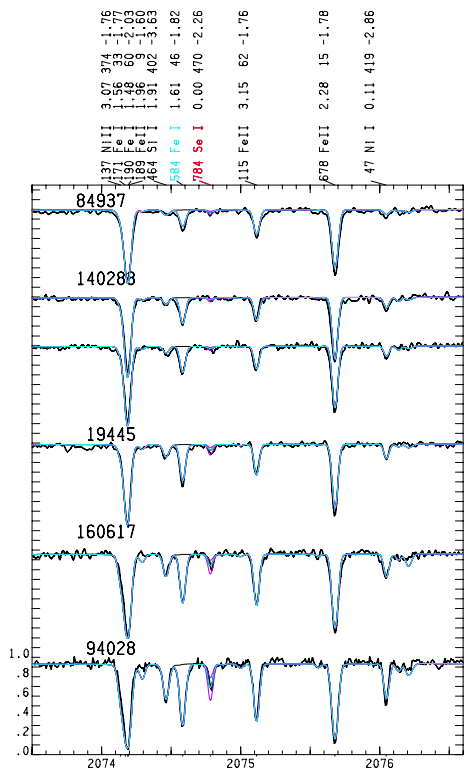}}
%\resizebox{\hsize}{!}{\includegraphics{figb207.eps}}
\caption{Spectra in the region of the \ion{Se}{I} 2074.794 {\rm \AA} line.}
\label {fig7}
\end{figure}

\begin{figure}
%\resizebox{11.0cm}{12.0cm}           	
{\includegraphics [scale=0.40]{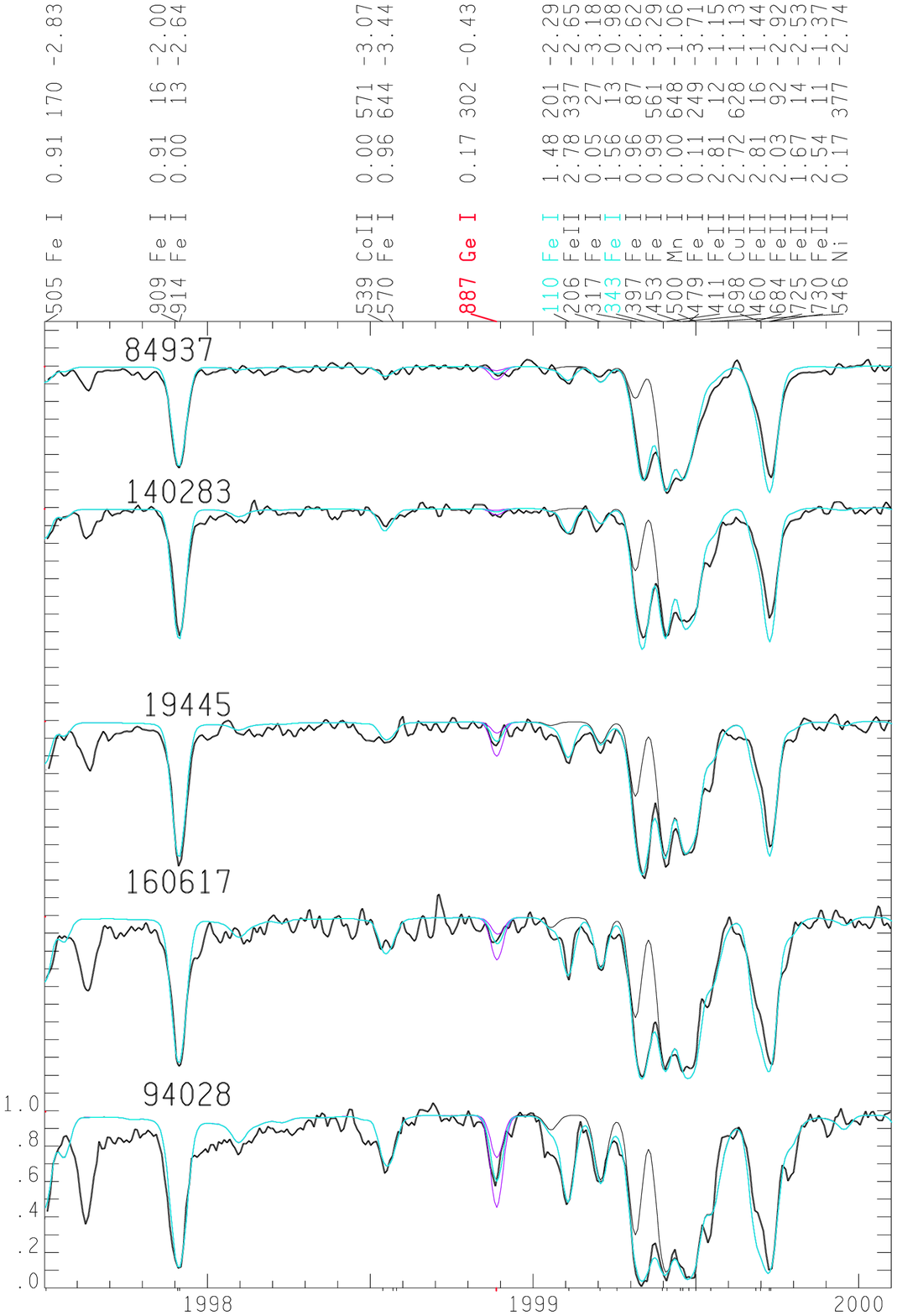}}
%\resizebox{\hsize}{!}{\includegraphics{figb199.eps}}
\caption{Spectra in the region of the \ion{Ge}{I} 1998.887  {\rm \AA} line.}
\label {fig8}
\end{figure}

\begin{figure}
%\resizebox{8.0cm}{12.0cm}           	
{\includegraphics [scale=0.40]{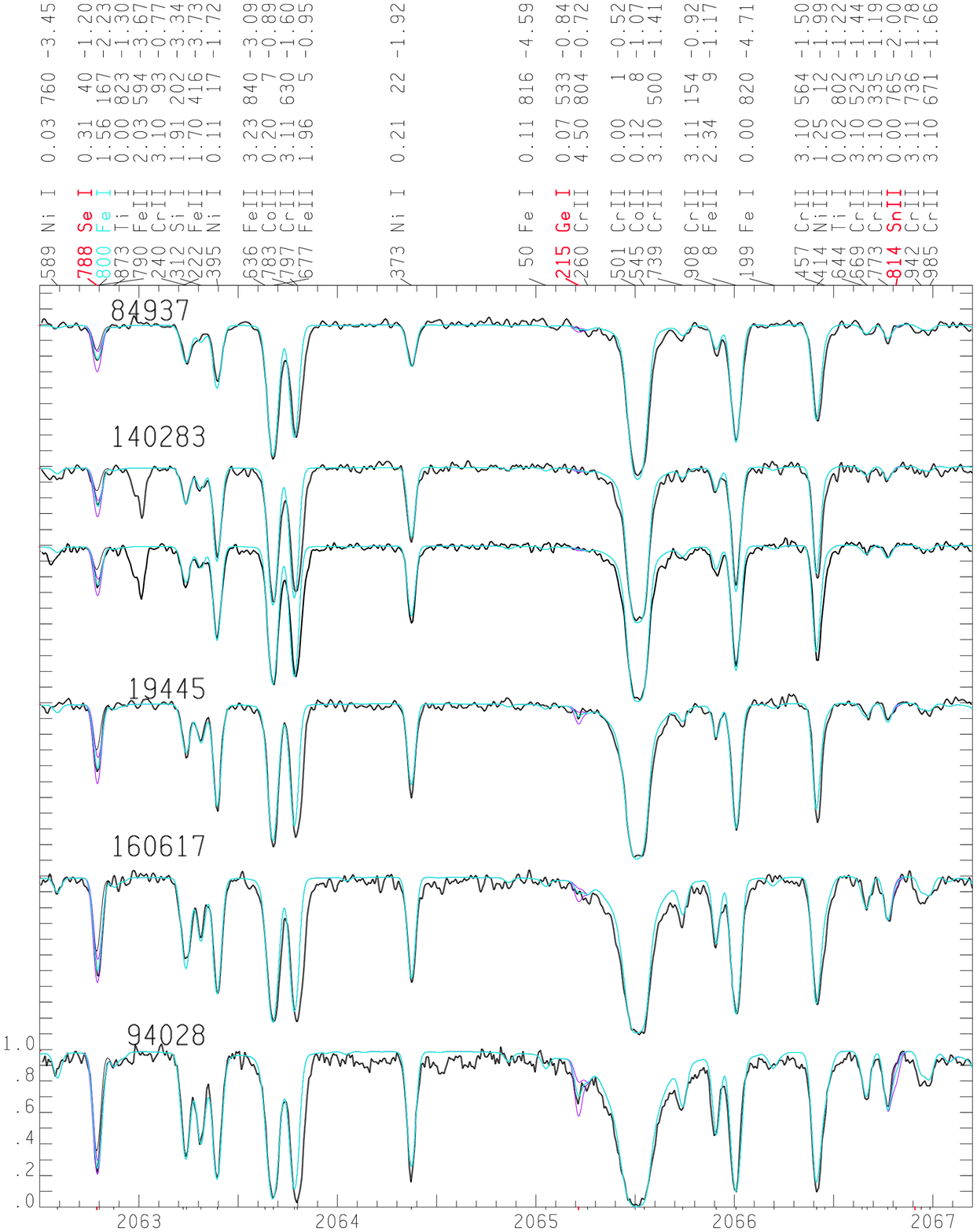}}
%\resizebox{\hsize}{!}{\includegraphics{figb206.eps}}
\caption{Spectra near the  \ion{Ge}{I} 2065.215 {\rm \AA} line. 
Here and in Figs.\ \ref{fig7}, \ref{fig10}, \ref{fig14}, and \ref{fig15}, 
the GO-7348 spectrum for HD\,140283 appears above that of GO-14672.}
\label{fig9}
\end{figure}

\begin{figure}
%\resizebox{12.0cm}{12.0cm}           	
{\includegraphics [scale=0.40]{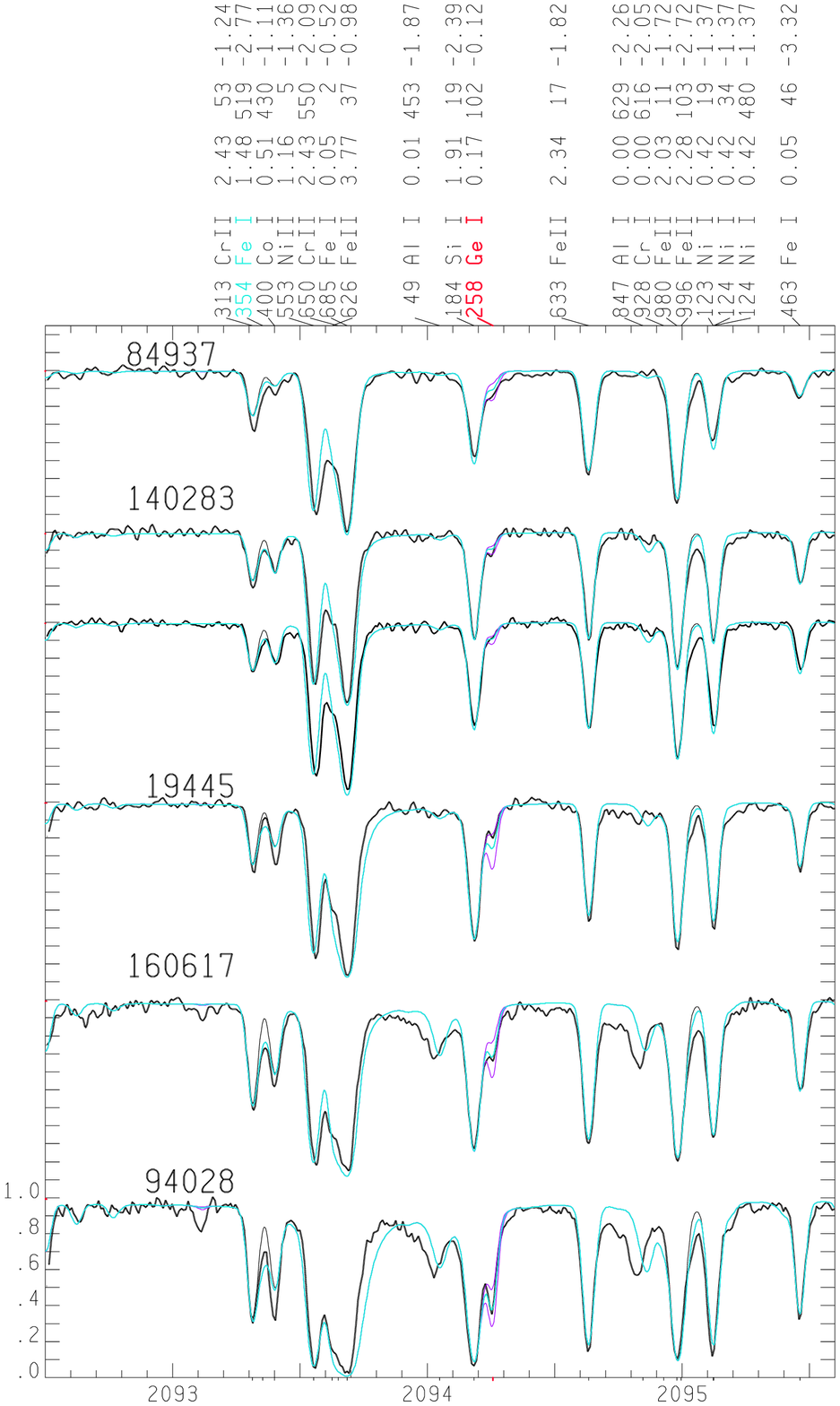}}
%\resizebox{\hsize}{!}{\includegraphics{figb209.eps}}
\caption{Spectra in the region of the  \ion{Ge}{I} 2094.258 {\rm \AA} line.}
\label{fig10}
\end{figure}

\begin{figure}
%\resizebox{8.0cm}{12.0cm}           	
{\includegraphics [scale=0.40]{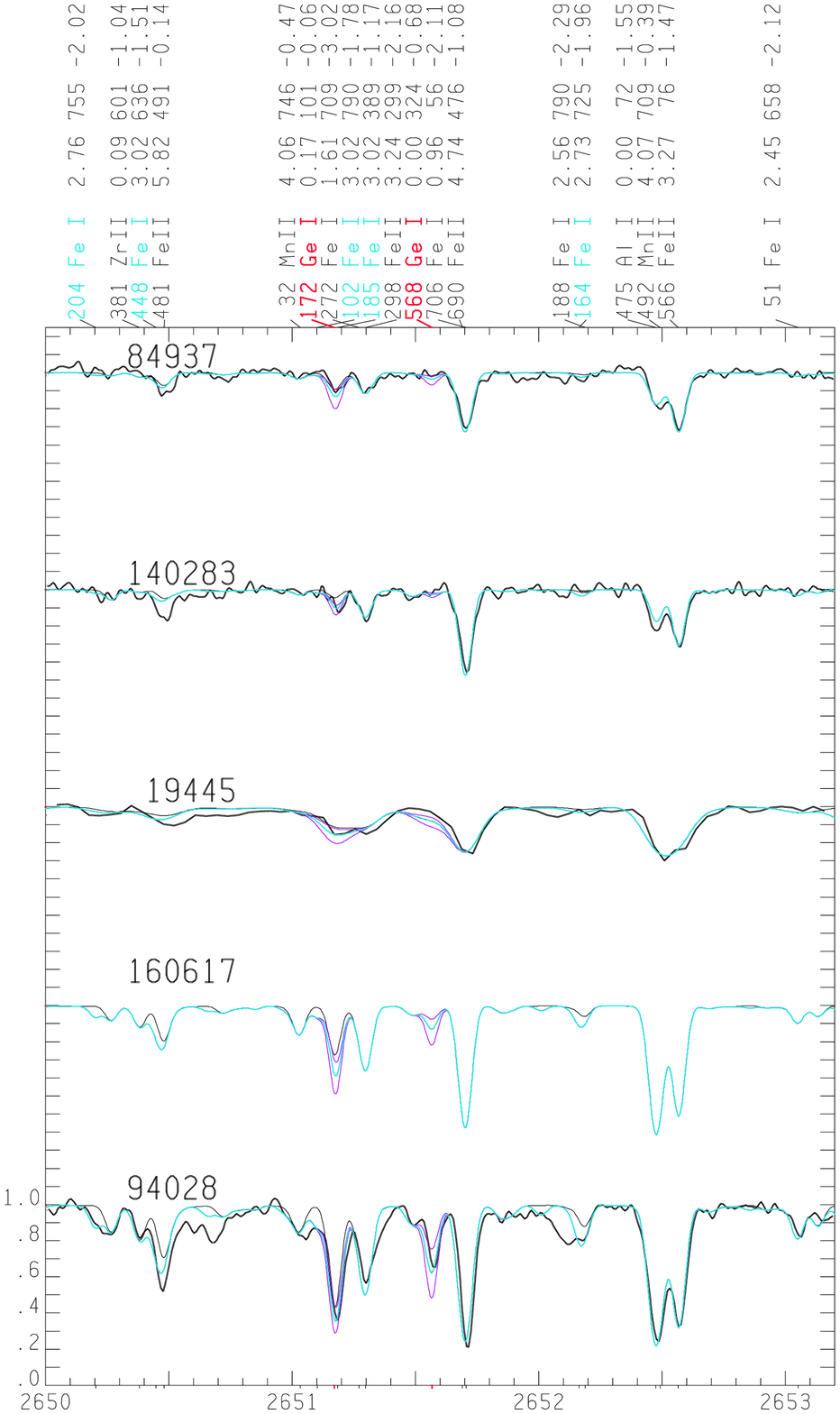}}
%\resizebox{\hsize}{!}{\includegraphics{figb265.eps}}
\caption{Spectra in the region of the \ion{Ge}{I} 2651.568 {\rm \AA} line. In this region and those of Figs.\ \ref{fig12} and \ref{fig13}, data for HD 160617 are lacking, and are from a lower-resolution E230M spectrum for 19445.  
The effect of weak newly identified and unidentified {\ion{Fe}{I}} lines 
is evident in the three E230H spectra; their effect persists but is difficult to discern at the E230M resolution.
}
\label{fig11}
\end{figure}

\begin{figure}
%\resizebox{10.0cm}{12.0cm}           	
%\includegraphics [scale=0.45]{figb270.eps}
{\includegraphics [scale=0.40]{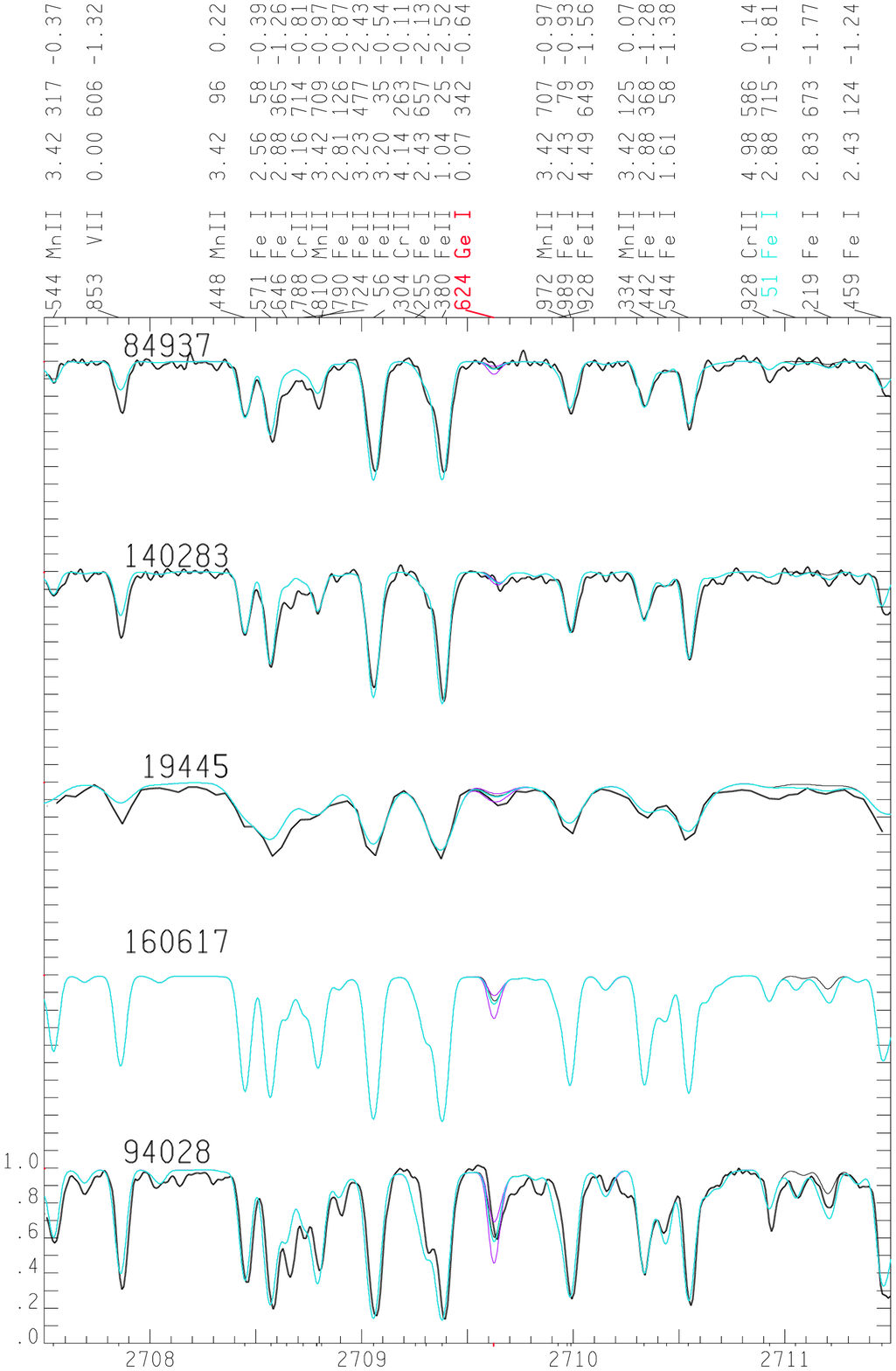}}
%\resizebox{\hsize}{!}{\includegraphics{figb270.eps}}
\caption{Spectra in the region of the \ion{Ge}{I} 2709.624 {\rm \AA} line. Unidentified lines are still present but less prominent 
here than in Fig.\ \ref{fig11}.}
\label{fig12}
\end{figure}

\begin{figure}
%\resizebox{9.0cm}{12.0cm}           	
{\includegraphics [scale=0.40]{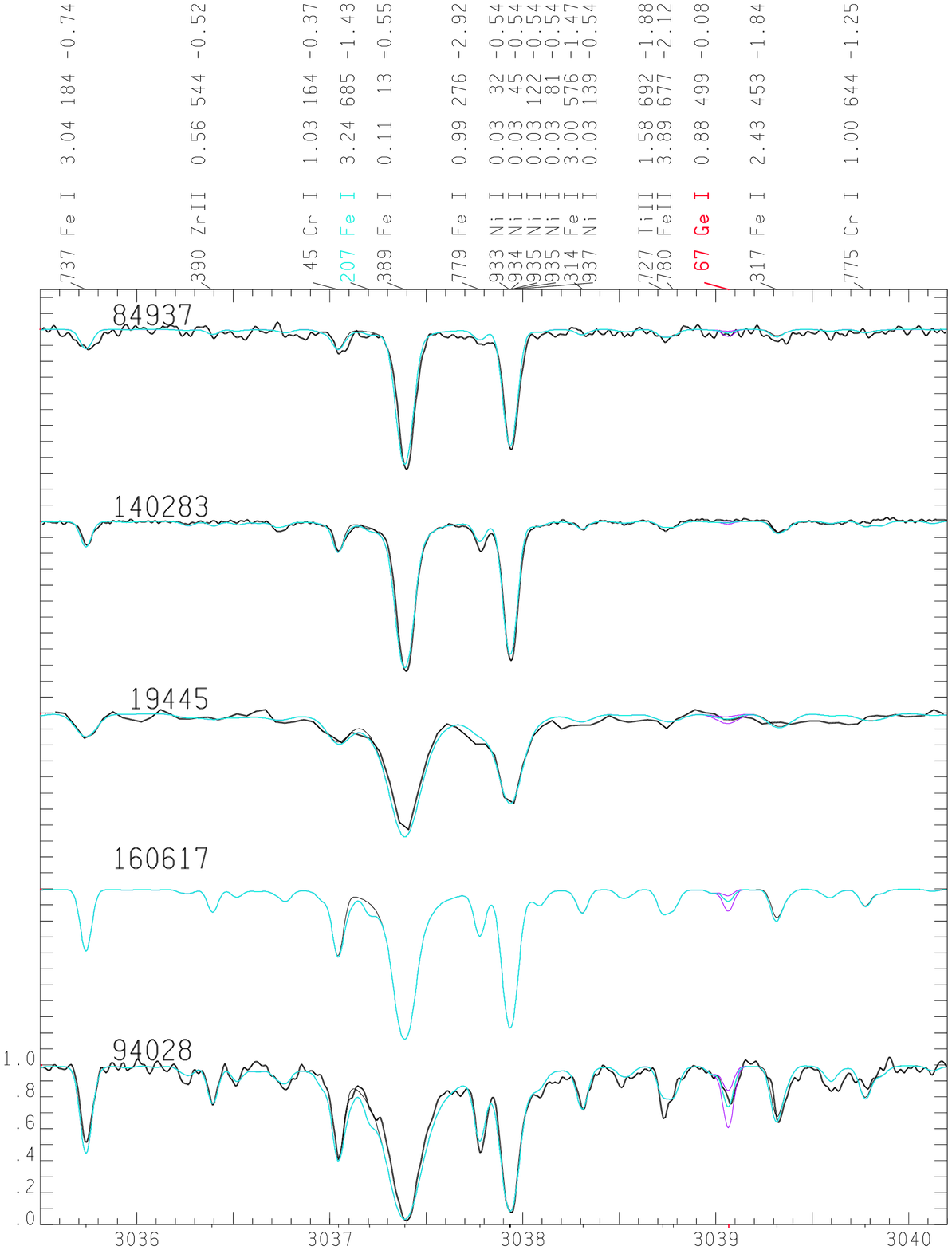}}
%\resizebox{\hsize}{!}{\includegraphics{figb303.eps}}
\caption{Spectra in the region of the \ion{Ge}{I} 3039.067 {\rm \AA}  line. Unidentified lines are virtually absent even in the 
HD 94028 spectrum.}
\label {fig13}
\end{figure}

\begin{figure}
%\resizebox{9.0cm}{12.0cm}           	
{\includegraphics [scale=0.40]{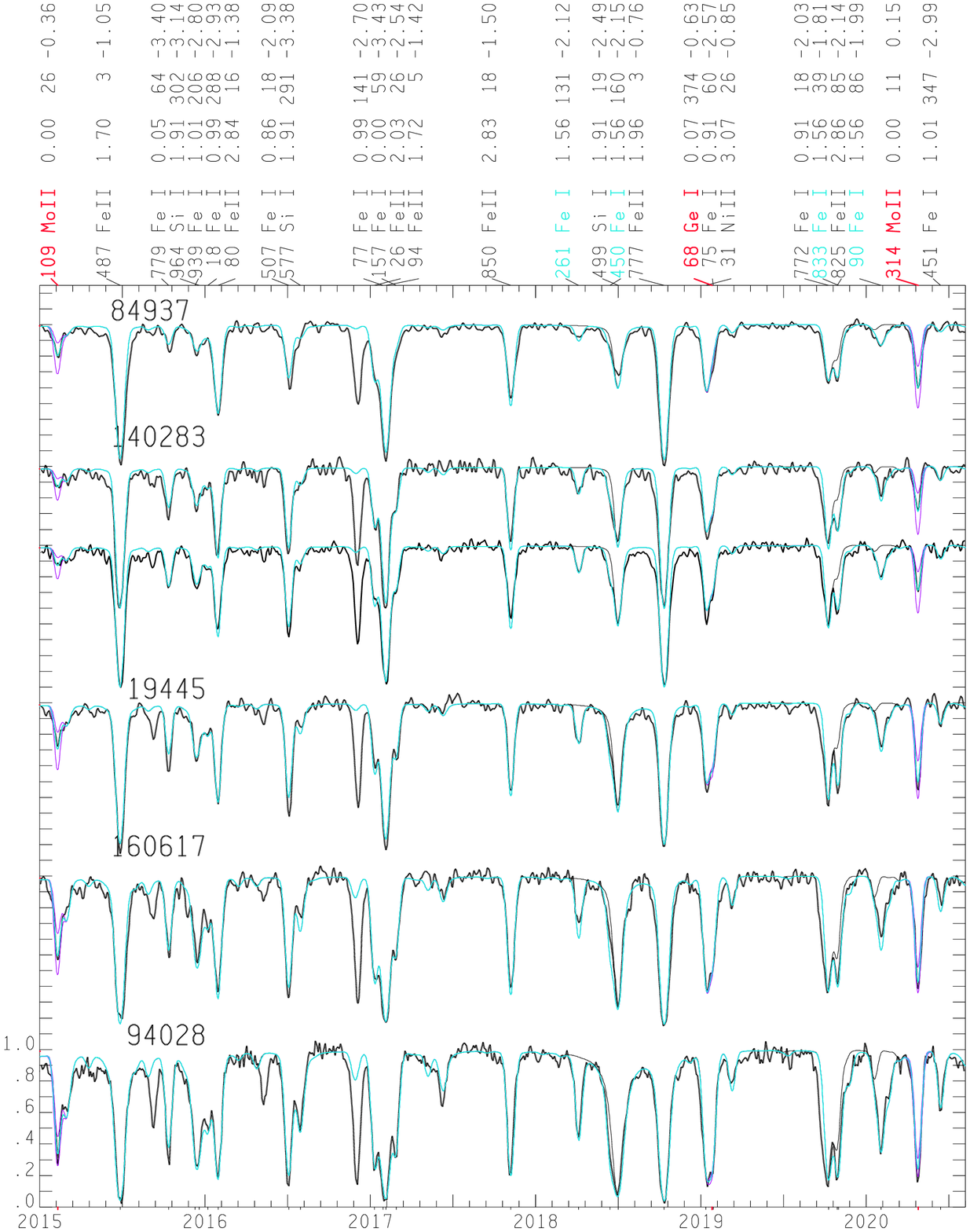}}
%\resizebox{\hsize}{!}{\includegraphics{figb201.eps}}
\caption{Spectra covering the \ion{Mo}{II} 2015.123 {\rm \AA} and
  2020.324 {\rm \AA}
  lines.}
\label {fig14}
\end{figure}

\begin{figure}
%\resizebox{9.0cm}{12.0cm}              
{\includegraphics [scale=0.40]{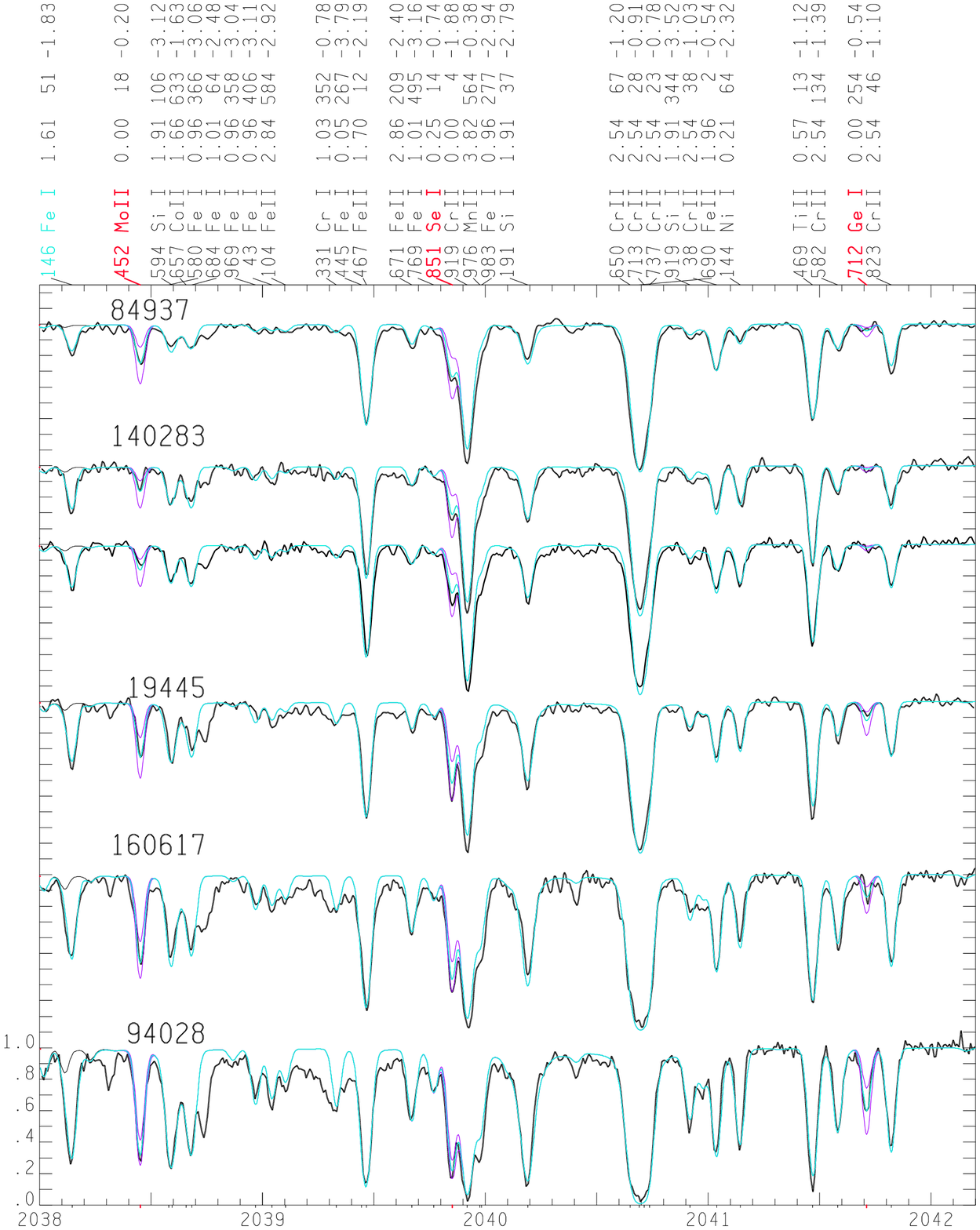}}
%\resizebox{\hsize}{!}{\includegraphics{figb203.eps}}
\caption{Spectra in the region of the
\ion{Mo}{II} 2038.452 {\rm \AA}, \ion{Se}{I} 2039.842 {\rm \AA}
  and \ion{Ge}{I} 2041.712 {\rm \AA}
   lines.}
\label {fig15}
\end{figure}

Calculations by \citet{AndrievskySK07} for the Na D lines in 
warm turnoff stars found an 
NLTE correction of $-0.3$ dex at [Fe/H] = $-2.5$, decreasing to <$-0.1$ dex
at [Fe/H] = $-3.5$. In general, the weaker the line the weaker the NLTE 
effect, as is consistent with weak-line formation at greater atmospheric depth 
and pressure. Their Fig.\ 4 shows that the Na D 5890\AA\ lines in 
high-resolution spectra of stars with these parameters 
have residual intensities of 0.75 to 0.85. 
Our UV trans-Fe resonance lines are rarely this strong, 
but are formed at shallower levels (Sec.\ \ref{meth}).
Nonetheless, in Figs.\ \ref{fig6} and \ref{fig7}  
the same LTE selenium abundance matches
each star's Se I profile of both the strong {\ion{Se}{I}} line at 1960.893\AA\
and the weak {\ion{Se}{I}} line at 2074.794\AA. 

Concerning the ionization balance, the {\tt SYNTHE} result for the abundance of the
minority neutral species agrees with that of the ionized species for elements
such as Fe and Ti that are represented by many lines of each species.
Furthermore, with the adoption in {\tt SYNTHE} of l/H = 0.5 models for the
weakest-lined stars (Sec.\ \ref{meth}), no dependence is seen of the abundance within
a neutral species on the lower excitation potential of the line, 
neither in the UV nor in the optical. 

%\clearpage
%

\subsection{Abundance patterns in the most metal-poor stars: HD\,19445, HD\,84937, HD\,140283, and HD\,160617}

With these results, we turn our attention to the abundance patterns exhibited within each star 
by the light trans-Fe elements and their heavier counterparts. 
We first examine the four most metal-poor stars analyzed here, those with metallicities 
approaching 1/100 solar: $\rm[Fe/H] \leq -1.8$. Their r-process content is judged from 
the [Eu/Fe] values of Table \ref{totheavy}.  With [Eu/Fe] = $-0.22$,
HD\,140283 has a less-than-solar proportion of r-process elements; 
the other three stars have modest enhancements, +0.37 $\leq$ [Eu/Fe] $\leq$ +0.51.

%\clearpage

%Figure \ref{pat4} 
%Next we compare the abundance pattern observed in 
%each of these four stars with the abundance pattern of calculations of different processes
%that can enrich the matter in heavy elements in the early Galaxy.  
The main r-process abundance pattern has been computed by 
\citet{Wanajo07} for hot and cold models. 
This theoretical pattern is known to give a good representation of
the abundance pattern of stars with extreme r-process enhancements [Eu/Fe] > 1, 
such as CS\,31082-001 \citep{HillPC02,BarbuySH11,SiqueiraSB13} or CS\,22892-052
\citep{SnedenMWP96,SnedenCL03}. 
But in metal-poor giants with more modest r-process enhancements, 
an overabundance is often observed of Sr, Y, and Zr, 
the first-peak elements with optical lines. 
As the r-process enhancement drops towards the solar value, 
some stars show elevated Sr abundances, with [Sr/Eu] 
ranging from the solar value to [Sr/Eu] > 1.
\citet{SiqueiraHB14}, \citet{RoedererKP16}, \citet{AokiIA17},  and \citet{SpiteSB18} 
see this in giants within our Galaxy, as does \citet{Roederer17} in giants in ultrafaint galaxies.

\clearpage

\begin{figure*}
\begin{center}
\resizebox{7.0cm}{!} 
%{\incluegraphics [clip=true]{pat4HD160617.pdf}}
%\resizebox{7.0cm}{!} 
%{\includegraphics [clip=true]{pat4HD84937.pdf}}
%\resizebox{7.0cm}{!} 
%{\includegraphics [clip=true]{pat4HD19445.pdf}}
%\resizebox{7.0cm}{!} 
%{\includegraphics [clip=true]{pat4HD140283.pdf}}
{\includegraphics [clip=true]{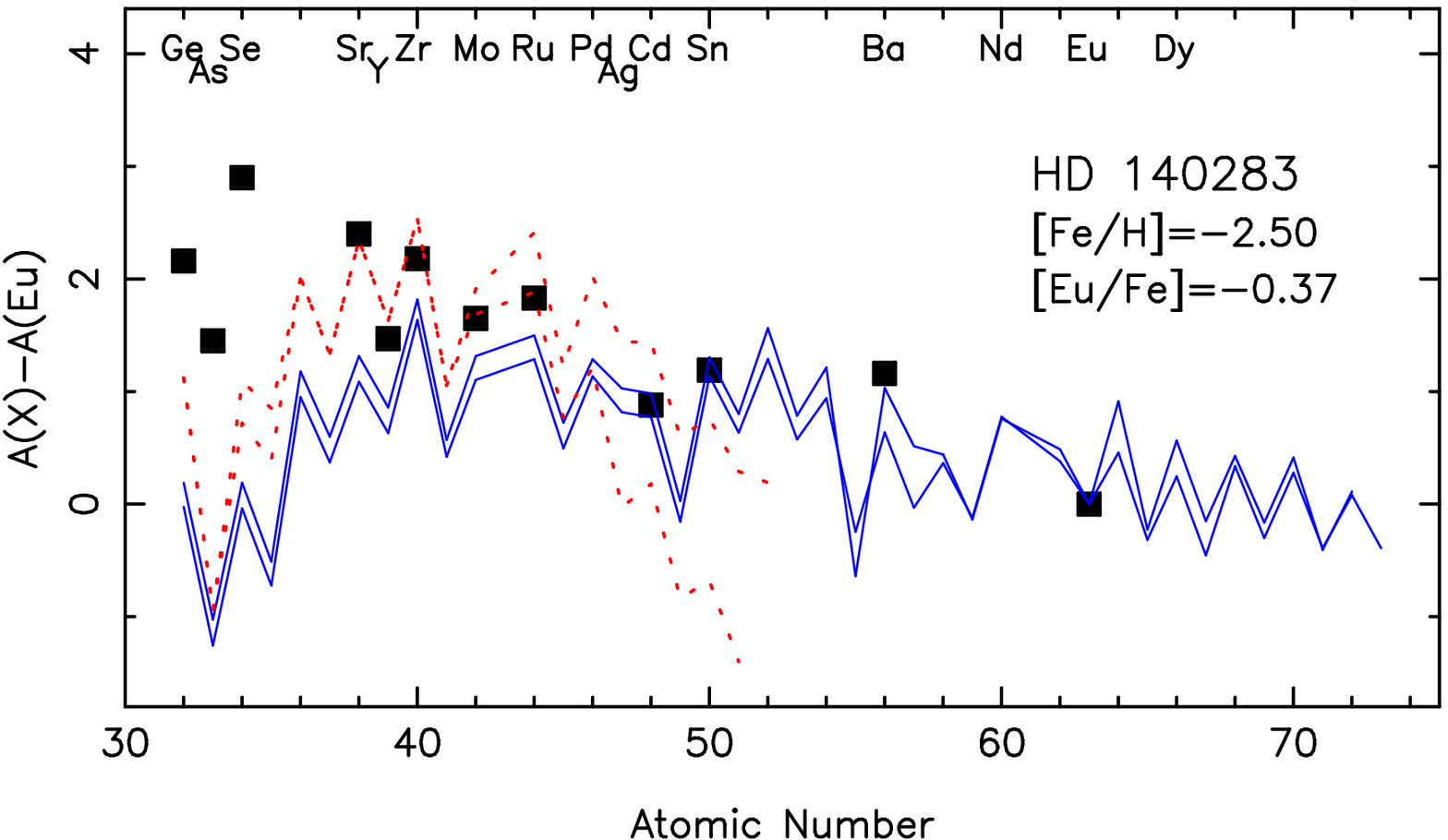}}
\resizebox{7.0cm}{!} 
{\includegraphics [clip=true]{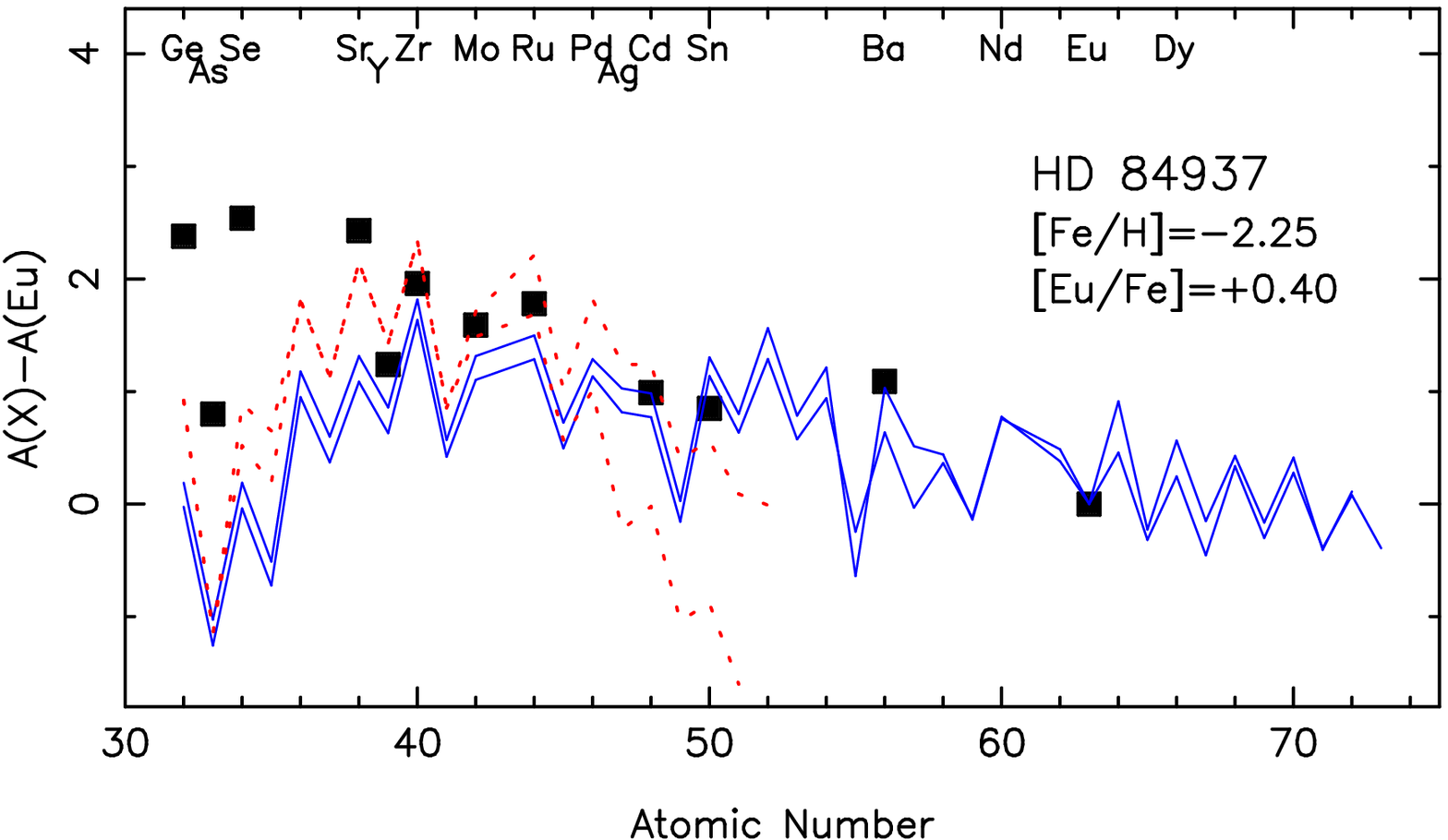}}
\resizebox{7.0cm}{!} 
{\includegraphics [clip=true]{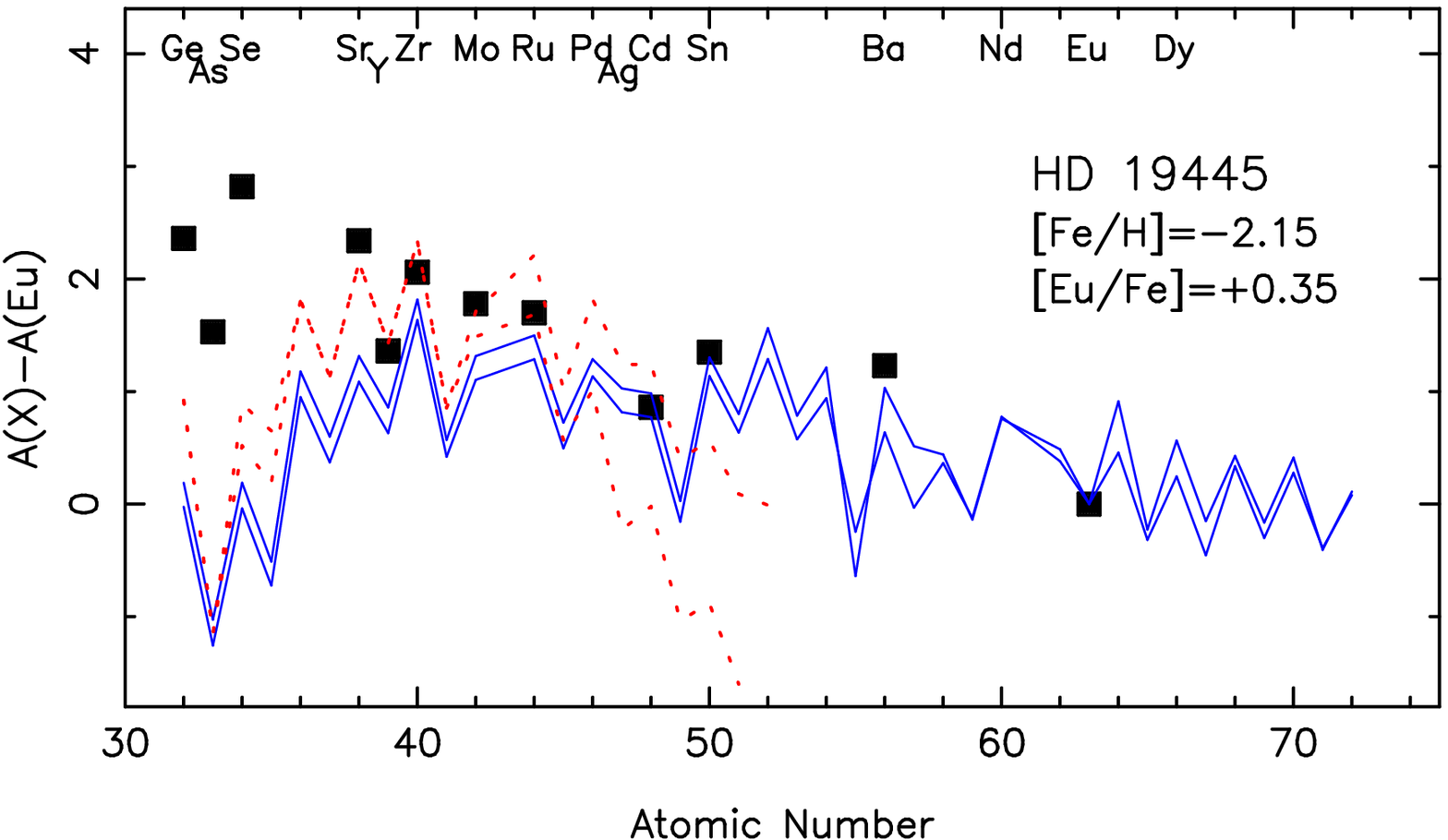}}
\resizebox{7.0cm}{!} 
{\includegraphics [clip=true]{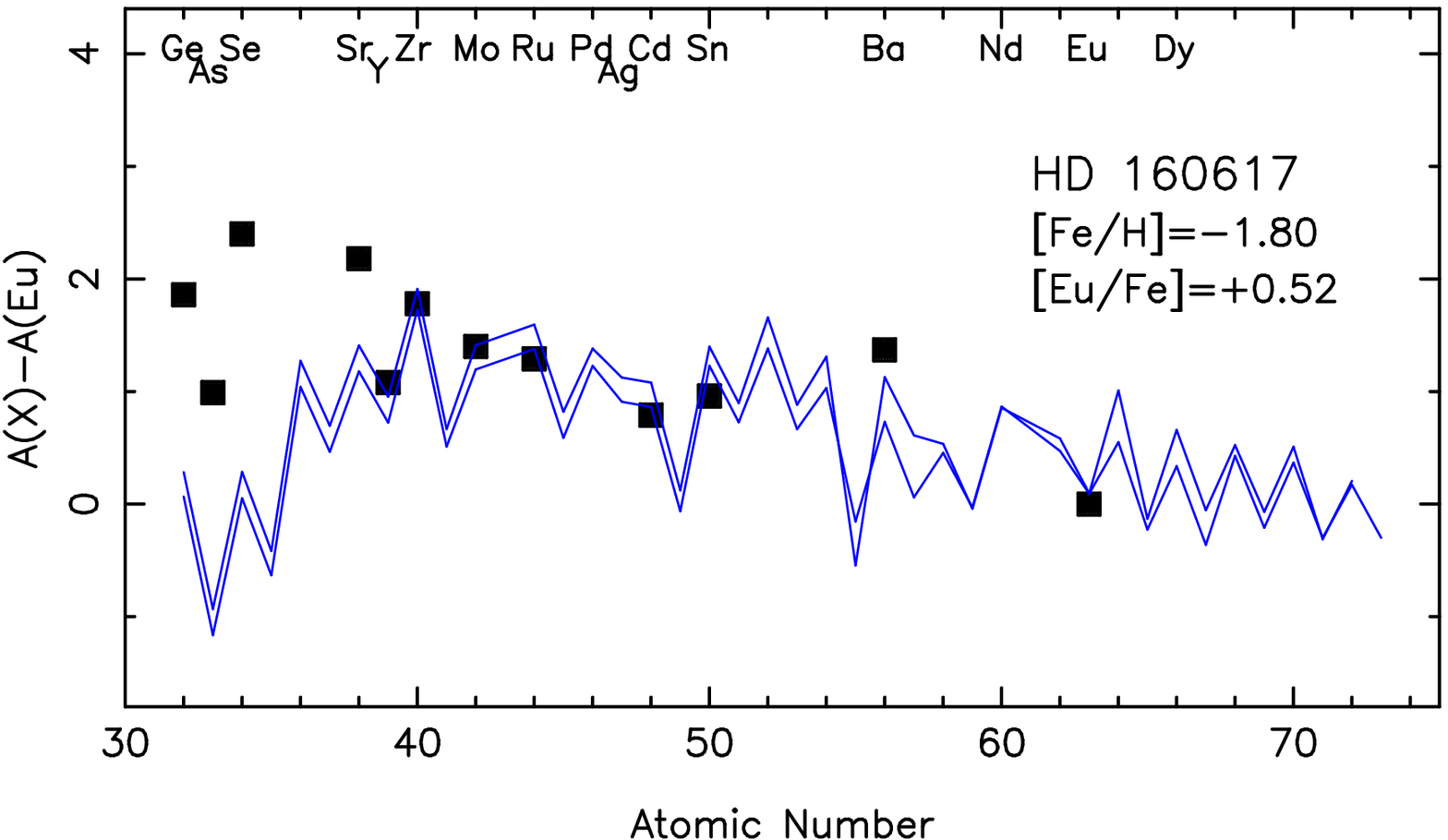}}
\end{center}
\caption[]{%The pattern of n
Neutron-capture abundances observed in the four sample stars with 
[Fe/H] $\leq$ -1.8 (black squares) are compared to 
the predictions of the main r-process following \citet{Wanajo07} (hot and cold models; blue lines) 
and the predictions of the weak-r process \citep{Wanajo13} with Proto-Neutron-Star masses M = 1.2 and 1.4 M$_{\odot}$ (red dotted lines).}
\label{pat4}
\end{figure*}

In Figure \ref{pat4}, we plot as a function of atomic number Z
the abundance pattern observed in each of these four stars (black squares) 
versus the \citet{Wanajo07} calculations (blue lines), 
both normalized to the abundance of Eu.  
Like the metal-poor giants of modest r-process enhancement, 
the four field halo turnoff stars of 
Fig.\ \ref{pat4} show Sr-Y-Zr abundances that are somewhat higher than the 
\citet{Wanajo07} predictions. 
The discrepancy is smallest for Zr and largest for Sr, 
and for all three elements the discrepancy increases as metallicity declines. 
In the three stars of lowest metallicity, the discrepancy extends to Mo and Ru as well.

All four stars show good agreement with main r-process predictions at the heavy end, for Ba and Eu. Their 
[Ba/Eu] ratios are always compatible with the main r-process, as expected when an 
s-process contribution is absent.
  More surprisingly, the ratios [Sn/Eu] and [Cd/Eu] appear to be also
  compatible with a pure main r-process. These elements are rarely observed in giants.

Other processes, collectively designated the "weak" r-process, 
can bring about an extra first-peak enrichment \citep{Wanajo13,CowanSL19}. %\\
Recently, \citet{AokiIA17} showed that small variations of the 
electron-capture supernovae parameters such as electron fraction
and Proto-Neutron-Star (PNS) mass 
could induce variations of the ejecta of the weak-r process that
can approximately reproduce the abundance patterns of the first peak
elements from Sr to Ba in their sample of giant stars. %\\

In Fig.\ \ref{pat4} we also compare the abundance pattern of our
most metal-poor stars to predictions (dotted red lines) of the weak-r process
by \citet{Wanajo13}, computed for PNS masses
of 1.2 to 1.4 M$_{\odot}$.
The Sr-Y-Zr-Mo-Ru 
abundances of these three stars 
are a much better match to these calculations 
than to the \citet{Wanajo07}
weak-r process contributions shown by blue lines in Fig. \ref{pat4}. 
However, the \citet{Wanajo13} calculations are not able to account for the high 
abundance of the lightest trans-Fe elements Ge, As, Se in any of 
the four stars.
A peculiarity of the weak-r process  \citep[following][]{Wanajo13} 
is indeed a rapid drop of the abundances below Z=40, 
which is not observed in these stars.

%FIG 16

\subsection{Abundance patterns in the less metal-poor star: HD\,94028}

\begin{figure}
\begin{center}
\resizebox{7.0cm}{!}
%{\includegraphics [clip=true]{pat4HD94028.pdf}}
{\includegraphics [clip=true]{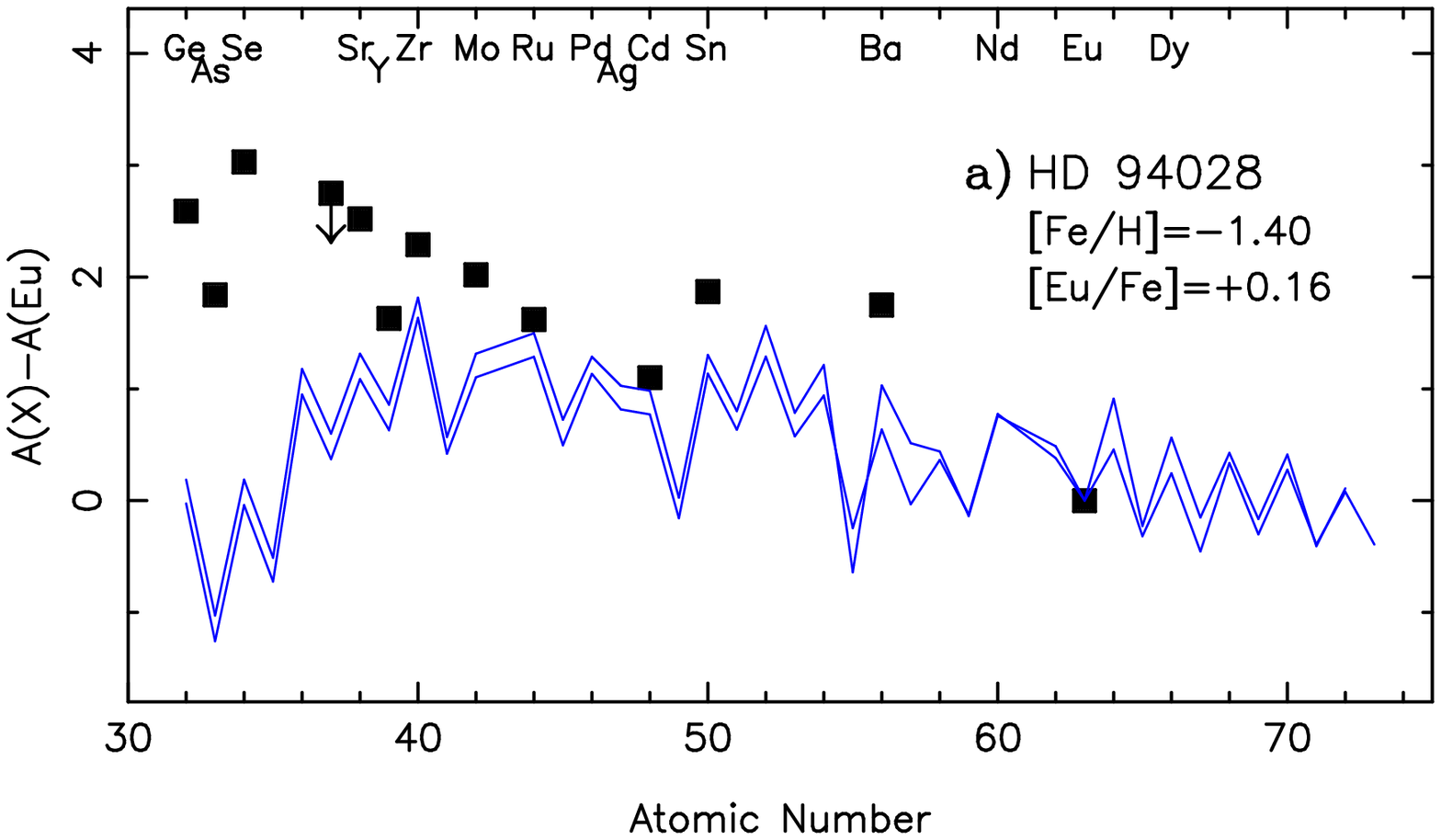}}
\resizebox{7.0cm}{!}
%{\includegraphics [clip=true]{pat4bHD94028.pdf}}
{\includegraphics [clip=true]{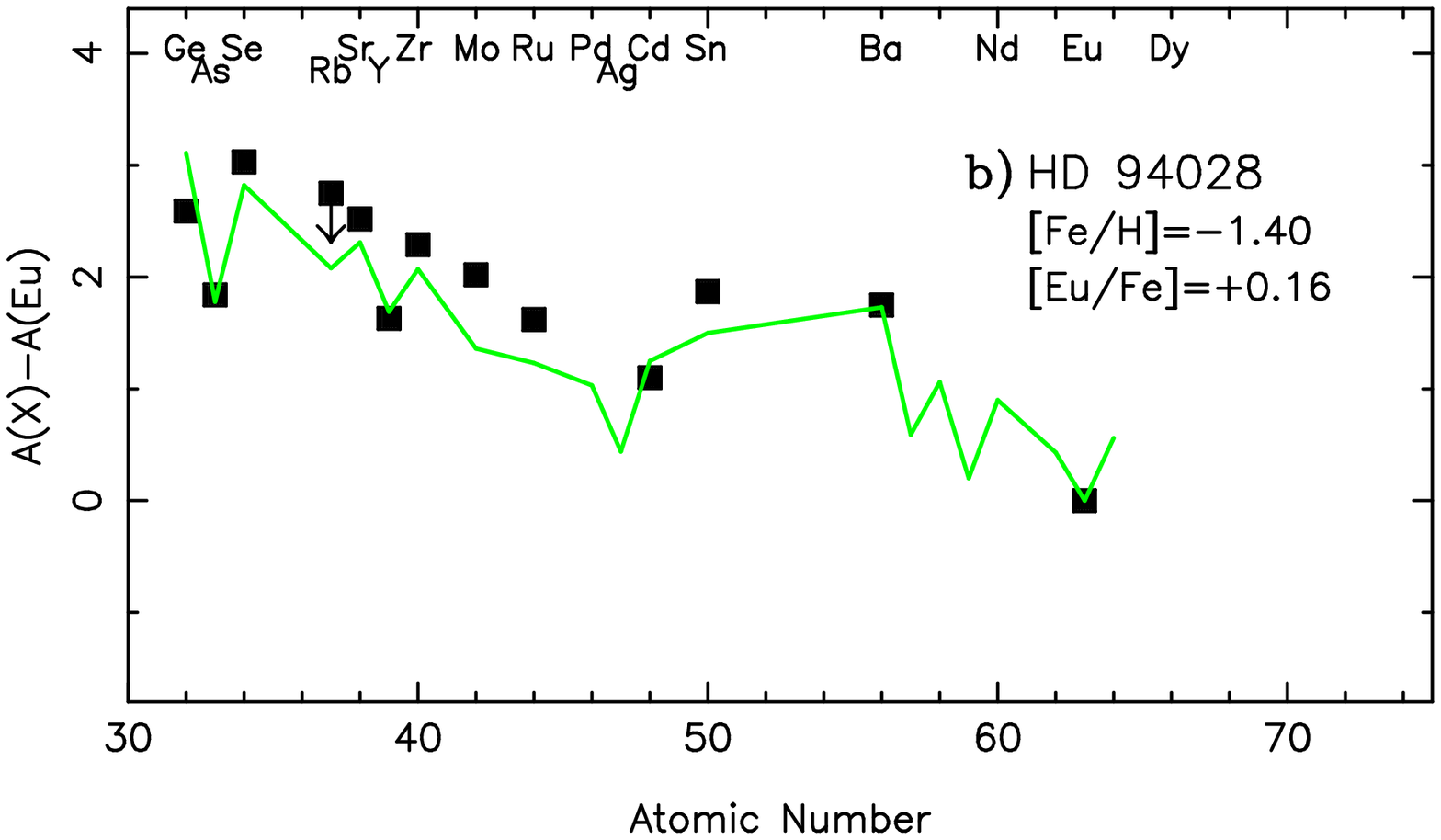}}
\resizebox{7.0cm}{!}
%{\includegraphics [clip=true]{pat4bHD94028.pdf}}
{\includegraphics [clip=true]{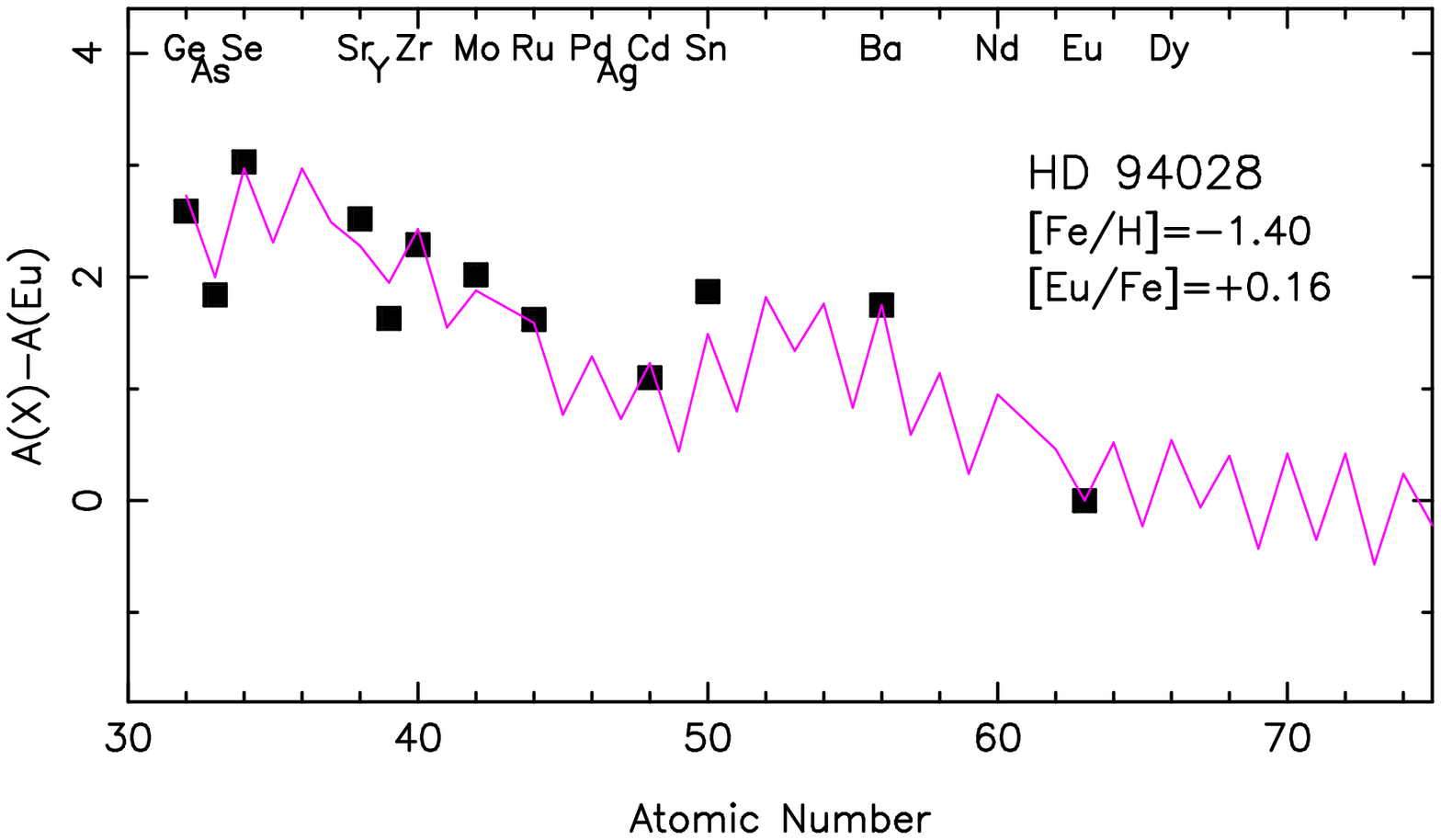}}
\end{center}
\caption[]{%Pattern of neutron-capture elements abundance in 
  Neutron-capture abundances observed in HD\,94028 are compared to 
  the predictions of the main r-process following \citet{Wanajo07}
  (hot and cold models, blue lines) in the top panel; 
  the solar abundance pattern (green line) in the middle panel; and 
  the predictions of the combined s, modified r, and i-processes 
  as computed by \citet{RoedererKP16} (their Figs.\ 6 and 7; red lines) in the bottom panel.
}
\label{pat4-2}
\end{figure}

With [Fe/H]=--1.4,  HD\,94028 is the least metal-poor star of our sample; it has a 
low r-process enhancement, [Eu/Fe] = +0.15. 
As seen in the upper panel of Fig.\ \ref{pat4-2}, the abundance pattern of this star is very different from the \citet{Wanajo07} main r-process pattern 
(blue lines). It also differs 
from the pattern behavior of the four more metal poor stars in Fig.\ \ref{pat4}. 
Not even the observed Ba/Eu ratio in Fig.\ \ref{pat4-2} fits the main r-process pattern, as \citet{RoedererKP16} also noted.

In the middle panel of Fig.\ \ref{pat4-2}, we compare the abundance pattern of HD\,94028 to the abundance pattern of the Sun. The agreement is much better. 
The ratio Ba/Eu is the same as observed in the Sun, which suggests the presence of s-process products
to the same degree. 
The light trans-Fe elements from Ge to Cd 
have abundances that remain higher than their solar proportions, but 
the excess is moderate.

In the lower panel of Fig.\ \ref{pat4-2}, we compare the abundance pattern of HD\,94028 to 
a composite template constructed by \citet{RoedererKP16}. 
To explain the observational distribution of element abundances in HD 94028, and in particular the high abundances they derive for Ge, As, and Se in this star, 
\citet{RoedererKP16} invoked contributions from three distinct 
mechanisms: the s-process, 
an r-process modified from that of the Sun, and 
an i-process, which designates neutron capture with a neutron flux intermediate between that of the s- and r-processes. 
Such i-process conditions 
may be found in multiple stellar sites, with both short and long time scales. 

\citet{RoedererKP16} followed \citet{CowanRose77} in investigating the i-process in an evolving star, when a rapid ingestion of a large quantity of H into the He-burning convective regions occurs, producing neutrons through the $^{13}$C($\alpha$,n)$^{16}$O reaction. 
\citet{RoedererKP16} 
computed the ejecta of this i-process at a metallicity [Fe/H]=--2, using the i-process trajectory of \citet{BertolliHP13} adjusted to maximize production in the As-Mo region. 
Their Fig.\ 7 shows a reasonably good match to the abundance distribution they derived for HD\,94028. Ours shows a similar fit, 
despite the Ge and As abundance discrepancies noted in Sec.\ \ref{compaRoed}. 

\citet{RoedererKP16} remarked that this i-process calculation
generally reproduced an observational trend they detected among HD\,94028 and the more 
metal-poor stars they had analyzed previously:
that [As/Ge] is generally super-solar while [Se/As] is sub-solar. 
At the same time they noted that Y was overproduced in their calculations,
and stated that their proposed detection of i-process products at early times is preliminary, requiring 
more comprehensive models.

We concur that more work needs to be done here. 
While we do find a super-solar ratio [As/Ge] for all 
these very metal-poor stars except HD\,84937, we also find  
the ratio [Se/As] to be super-solar in all our stars (Table \ref{totheavy}). 
We are also concerned that the \citet{RoedererKP16} Fig.\ 7 comparison 
rests on a substantial r-process modification. 
Prior to adding their i-process contribution to their template, 
they added to the solar r-process abundances 
the differences from the \citet{SimmererSC04}
solar r-process distribution that \citet{RoedererLS12} 
and \citet{RoedererSL14} had derived 
for the star HD 108317, a very metal-poor giant of somewhat 
higher r-process content.
Based on the bottom panel of Fig.\ 9 of \citet{RoedererSL14}, this 
increased the amount attributed to the r-process 
by +1.2 dex for Ge and +0.7 dex for As. 
Whether such a modified r-process has similarly contributed 
to the abundance distribution of trans-Fe elements of some metal-poor stars but 
not to that of the Sun remains an open question.%}

\subsection{The odd-even effect and its metallicity dependence among trans-iron elements in metal-poor stars}
\label{SrYZr}

In Figs.\ \ref{pat4} and \ref{pat4-2}, 
all five metal-poor stars exhibit an 
odd-even effect in the trans-Fe elemental trios Ge-As-Se and 
Sr-Y-Zr: the abundances are depressed of the odd-Z elements 
As and Y relative to those of the adjacent even-Z elements. 
Among such unevolved metal-poor stars generally, an odd-even 
effect is also seen in the sodium and aluminum abundances. 
Its size increases as metallicity declines, due to 
the metallicity dependence of the neutron seed 
in the progenitor of explosive synthesis. 

For the trans-Fe elements, 
a hint of such a metallicity dependence is suggested, 
in that the size of the odd-even effect in 
both Ge-As-Se and Sr-Y-Zr is larger in the most metal-poor stars 
of Fig.\ \ref{pat4} than in the most metal-rich star HD 94028 
in Fig.\ \ref{pat4-2}. However, conclusive evidence depends on a 
larger sample of stars spanning a wider metallicity range.

As noted above, extensive Sr, Y, and Zr abundances for metal-poor stars have been obtained 
from ground-based optical abundances. 
Consequently we have examined the behavior of the Sr-Y-Zr abundances from two such 
studies with larger samples, those by \citet{Peterson13} in 29 more metal-rich dwarfs, 
and by \citet{SpiteSB18} in 14 very metal-poor giants. Subtracting the 
average of [Sr/Fe] and [Zr/Fe] from the [Y/Fe] value for each star in each study 
revealed that the Y abundance is once again generally suppressed with respect to their 
average.

A plot versus metallicity of these values for both groups provided a fit to all points that shows a 
strong correlation, with a correlation coefficient $R$ = 0.82, 
and an individual standard deviation of 0.082 dex for the giants and 0.067 dex for the dwarfs.
At the upper metallicity limit of the dwarfs, [Fe/H] = -1.2, [Y/Fe] - <[Sr/Fe]+[Zr/Fe]> = 0.0, 
indicating the Y abundance is not depressed below its solar value 
relative to the average abundance of the adjacent elements Sr and Zr. 
The Y relative abundance does become depressed at lower metallicities, 
however. [Y/Fe] - <[Sr/Fe]+[Zr/Fe]> decreases by 
0.15 dex for every 1 dex decrease in [Fe/H], becoming -0.4 at [Fe/H] = -4, 
the lower metallicity limit of the giants. 

\subsection{i-process calculations and the odd-even effect in trans-iron elements}
\label{iprocoddeven}
 
Generally speaking, the \citet{Wanajo13} calculations (pink lines) in Fig.\ \ref{pat4} fail to reproduce the overall rise of the lightest trans-Fe elements in the more metal-poor stars, while the \citet{RoedererKP16} composite template (pink line) in the bottom panel of Fig.\ \ref{pat4-2} reproduces the overall distribution of HD 94028 quite well.
In contrast, despite having adopted [Fe/H] = -2
for the i-process calculations of HD 94028 (at [Fe/H] = -1.4),
the \citet{RoedererKP16} template in Fig.\ \ref{pat4-2}
underestimates the odd-even effect of its trans-Fe elements,
especially for Sr-Y-Zr; while the \citet{Wanajo13} calculations 
in Fig.\ \ref{pat4} 
match the odd-Z to even-Z trans-Fe abundance variations 
quite well for both Sr-Y-Zr and Ge-As-Se.

From this it seems that while the i-process offers promise 
to explain the excess of trans-Fe abundances in significant classes 
of metal-poor stars, more calculations are needed to judge whether 
its products are representative of the abundance distribution of the 
light trans-Fe elements in the majority of field halo stars with low to moderate 
enhancements of the r-process heavy elements. In particular, the presence 
of a trans-Fe odd-even effect suggests that more useful 
i-process calculations should include yields for model star progenitors that span 
the range of metallicities $-4 \leq$ [Fe/H] $\leq -1$. As noted 
in the Introduction, two sets of recent i-process calculations do this.

\citet{RitterHJ18} extended the NuGrid program results that 
provide stellar overproduction factors for virtually all elements 
synthesized in stars with initial masses 1 -- 25 M\sun. Their models 
include the evolution of each star before, during, and after 
each condition amenable to nucleosynthesis, incorporating the 
various neutron addition processes based on the associated neutron fluxes. 
In this work they included progenitor metallicities ranging from 
mildly metal-poor to extremely metal-poor halo stars, again including the 
nucleosynthetic yields of both massive stars and lower mass stars on the AGB. 

One suggestive result of these calculations is that where Sr-Y-Zr is copiously 
produced, the adjacent element rubidium (Rb; Z=37) is produced in even greater 
overabundance. To check this possibility, {\tt Turbospec} calculations were run 
near the {\ion{Rb}{I}} line at 7800.26\AA, adopting the \citet{Warner68a} value log gf 
= +0.137 dex. These were compared to a high-quality red spectrum of HD\,94028 
formed from the coaddition of two UVES spectra. As shown in 
%{\bf 
the upper panels of Fig.\ \ref{pat4-2}, 
this comparison led to an upper limit to the HD 94028 Rb abundance. 
This limit is not inconsistent with these predictions, but cannot confirm them. 
We also searched for this line in UVES archival spectra for HD 84937, HD 140283, and 
HD 160617, and in a spectrum obtained at the TBL (Pic du Midi) for HD 19445, 
but no hint of its detection was found.

Unfortunately, for most trans-Fe elements the \citet{RitterHJ18} yields 
are difficult to compare directly with observations, 
for they often exhibit significant variations between small changes in mass. 
This is not unexpected given the variety of conditions involved, and the 
approximations necessary to model them. As these assumptions improve, these 
calculations might provide a comprehensive benchmark for comparison against 
present and future observational constraints from metal-poor stars.

The \citet{BanerjeeQH17} calculations were done for metallicities from one-tenth 
to 1/10,000 solar, but only for a single mass of 25~M\sun. They are based on the 
ingesting varying amounts of protons into its He-burning shell. Their choice 
of mass stems from their arguments that synthesis from proton ingestion only 
occurs for stars of 20 -- 30~M\sun. They point out 
that the burning of the ingested protons is very difficult to model because
the convective mixing can only be approximated very simply. Despite these 
limitations and simplifications, their results do 
show a clear odd-even effect among the trans-Fe elements, one which 
increases as metallicity drops. This seems to be a very promising test case.

\subsection{Additional observational correlations among elemental abundances} 
\label{obscorr}

Among the abundances we derive for the five stars considered here, 
outside of the odd-even effect and the two exceptions noted just below, 
the excess or deficiency of a particular element shows no correlation with that of 
any other element, whether within or beyond the mass range $30<Z<52$.
In particular, none of these elements individually tracks the overall 
enhancement of r-process heavy elements among these stars. They do among 
metal-poor giants that are highly enhanced in r-process elements, but that 
is not the case among these dwarf stars, nor those of \citet{Peterson13}; 
with but a single exception, all of these have [Eu/Fe]$\leq$0.5. Four of the 
\citet{SpiteSB18} giants have high r-process content, +0.7 $\leq$ [Eu/Fe] 
$\leq$ +1.65, but show no obvious correlation for any trans-Fe elemental 
abundance, nor for the Sr-Y-Zr odd-even effect. For all four have 
-2.9 $\leq$ [Fe/H] $\leq$ -2.7 and -0.38 $\leq$ [Y/<Sr+Zr>] $\leq$ -0.14, 
straddling the best-fit line with rather large scatter.

The first exception is that even when molybdenum is highly enhanced, 
the heaviest trans-Fe elements are not. 
\cite{Peterson11} found Mo (Z=42) to be highly overabundant with respect to iron, 
but Cd (Z=48) and Sn (Z=50) were found at nearly their solar proportions.
This suggested the formation of trans-iron elements of lowest Z 
in the low-entropy regime of a high-entropy wind, 
which for certain parameters leads to excess production of 
primarily the 
trans-iron elements of lowest Z \citep{FarouqiKP09,FarouqiKM09,FarouqiKP10}. 
Such a specific constraint led \cite{Peterson11} 
to note that only a few individual supernovae must have contributed to 
the heavy-element content of these stars, even in HD~94028 at a metallicity 
[Fe/H] = --1.4, where the 
observed heavy-element abundance distribution usually requires
the accumulation of products from multiple events.

The second exception is that, with our addition of Ge abundances,  
we find that the germanium abundance does track that of molybdenum, but 
at a level that is dramatically lower than that of the solar ratio. 
Despite the wide star-to-star range 
of their individual abundances with respect to iron, 
[Mo/Ge] is essentially constant in all five stars at an average 
<[Mo/Ge]> = 1.18 dex, with an individual standard deviation of 0.10 dex. 
Among these five stars, this suggests that whatever has overproduced Mo
has underproduced Ge with respect to iron. This further constrains
both the mechanism
and the number of events contributing heavy elements to each star.

Unfortunately, neither the \citet{SpiteSB18} nor the \citet{Peterson13} 
results include germanium, because the single {\ion{Ge}{I}} line potentially 
observable from the ground falls at 3039\AA, where telluric absorption 
increases suddenly and dramatically toward the blue. Ground-based echelle 
spectra are usually very poorly exposed there, or cut off entirely just 
redward. Consequently this connection must await further 
observation and analysis for verification in the low-metallicity 
domain [Fe/H] $<$ -1.5.

To date, we have not found any calculations, i-process or otherwise,  
that predict a high, constant [Mo/Ge] ratio at these metallicities. 
One interpretation is that mentioned above, that the lightest trans-iron elements are 
preferentially formed in different processes, or perhaps in different environments 
of the same process, than are the heavier trans-iron elements. 
Regardless of the exact circumstances, 
this high [Mo/Ge] ratio in metal-poor dwarfs 
must return quite quickly to the solar ratio 
as iron abundance increases above [Fe/H] = -1.5. 
Consequently, despite the obscurity 
surrounding the mechanisms of the production of trans-Fe elements, 
our work nonetheless offers the promise that abundance 
determinations of molybdenum 
and germanium in the same stars as heavier trans-Fe elements 
might provide a long-sought way to 
identify the population from which a given star originated. 

%FIG 18

\section{Summary} 
We have obtained new high-resolution, high-signal to noise UV spectroscopic observations  employing the E230H grating of STIS onboard HST of the metal-poor turnoff stars \object{HD\,19445} and \object{HD\,140283} below 2000 {\rm \AA}. 
For these stars plus \object{HD\,84937}, \object{HD\,94028}, and \object{HD\,160617},  
we have analyzed similar UV spectra obtained for them previously  to determine abundances for the lightest heavy elements, Ge, As, Se, Mo, Cd, and Sn. Concurrent analysis of optical archival spectra have yielded the abundances of Sr, Y, and Zr, along with the heavy elements Ba and Eu.

Comparing the results from our two separate approaches indicates that the largest 
source of recognizable systematic error in the UV arises from the continuum placement in crowded regions. Comparisons with other studies indicate that the choice of trans-Fe lines for 
analysis and the choice of gf-values for those lines often contribute to systematic 
differences, once again affecting crowded regions the most. 

For germanium in particular, we show that 
the inclusion of newly identified {\ion{Fe}{I}} lines \citep{PetersonK15, PetersonKA17,PetersonKA20} 
aids in deconvolving the blending of weak lines and in continuum placement, 
but that as yet unidentified lines can still be 
influential, especially in the 2650 -- 2700\AA\ region. 

Again concerning germanium, our calculations that do 
incorporate newly identified {\ion{Fe}{I}} lines 
confirm that the \citet{LiNPW99} gf-values are on a consistent scale for the eight lines examined. 
These include the {\ion{Ge}{I}} 3039\AA\ line most often adopted in analyzing giants, which will enable 
results for dwarfs and giants to be placed on a more reliably consistent scale. Other 
elements, notably Mo, would also benefit from improved gf-value scales. 

From the reanalysis of optical archival spectra for these stars, we 
present abundances for the trans-Fe elements Sr, Y, and Zr, 
and the heavier elements Eu and Ba, which are largely due
to the r-process and s-process respectively.
Their inclusion enables us to examine patterns of light-Fe and heavy r-process elemental 
abundances together, to better constrain nucleosynthesis calculations and 
overall observational trends.

In contrast to the regular pattern exhibited by the r-process elements 
in metal-poor stars \citep{SnedenCG08}, 
the lighter elements studied here show a 
wide diversity of the excess or deficiency of a particular element versus iron. 
While this is already known, it remains subject to a range of interpretations.

The abundance of germanium with respect to that of iron ranges from 
the solar level to a deficiency of almost a factor of ten.
The strongest abundance 
excesses with respect to iron are seen in molybdenum, first 
noted by \cite{Peterson11}. Neither the germanium deficiency nor the
molybdenum excess is correlated with metallicity in this work, nor are such correlations 
generally seen in other studies of their respective abundances. 

Surprisingly, we find that these two elemental abundances are strongly
correlated with one another. Compared to the solar proportion,
molybdenum is an order of magnitude more
abundant than germanium in all five metal-poor stars studied here.

With measurements of a larger number of trans-iron elements, we uncover 
another unexpected result as well. The trans-iron elements exhibit an odd-even
effect, in both the Ge-As-Se and the Sr-Y-Zr elemental trios, in which
the element with odd Z shows a reduced abundance relative to its even-Z
neighbors. This tendency increases as metallicity decreases, as we show most 
clearly by evaluating Sr-Y-Zr measurements from our independent earlier optical studies of metal-poor stars.

\begin {acknowledgements}
We are indebted to Don Morton for the verification of oscillator strengths adopted in Table 3 for this work. We also thank Jim Lawler 
and the anonymous referee 
for their comments. Support for RCP for this work was provided by NASA through grant number GO-14762 from the Space Telescope Science Institute, which is operated by AURA, Inc., under NASA contract NAS 5-26555. MS is supported by the ``Programme National de Physique Stellaire'' and the ``Programme National de Cosmologie et Galaxies (PNCG)'' (CNRS-INSU).
BB acknowledges partial financial support from FAPESP, CNPq, and CAPES - Financial code 001.
\end{acknowledgements}

\bibliographystyle{aa}
{}

\end{document}